\documentclass[
    11pt,
    letterpaper,
    reprint,
    nofootinbib,
    notitlepage,
    superscriptaddress,
    aps,prx
]{revtex4-2}

\usepackage{amsmath,amssymb,amsfonts}
\usepackage{empheq}
\usepackage{physics}

\usepackage{amsthm}

\usepackage{subdepth} 

\usepackage{bm}
\renewcommand{\mathbf}{\bm}
\usepackage{dsfont}
\renewcommand{\mathbb}{\mathds}

\usepackage[svgnames,dvipsnames]{xcolor}
\usepackage{graphicx}
\definecolor{NewBlue}{rgb}{0.1, 0.1, 0.7}
\definecolor{NewRed}{rgb}{0.7, 0.1, 0.1}
\usepackage[colorlinks,
    linkcolor=Maroon,
    citecolor=NewBlue,
    urlcolor=NewRed]{hyperref}

\usepackage{cleveref}

\newcommand{\avg}{\ev}

\renewcommand{\t}[1]{\mathrm{#1}}

\renewcommand{\phi}{\varphi}

\newcommand{\LigoMIT}{LIGO Laboratory, Massachusetts Institute of Technology, Cambridge, MA 02139}
\newcommand{\MechMIT}{Department of Mechanical Engineering, Massachusetts Institute of Technology, Cambridge, MA 02139}

\begin{document}

\title{Exceptional-point Sensors Offer No Fundamental Signal-to-Noise Ratio Enhancement}

\author{Hudson Loughlin}
\email{hudsonl@mit.edu}
\affiliation{\LigoMIT}
\author{Vivishek Sudhir}
\affiliation{\LigoMIT}
\affiliation{\MechMIT}

\date{\today}

\begin{abstract}
    Exceptional-point (EP) sensors are characterized by a square-root resonant frequency bifurcation in response to 
    an external perturbation. This has lead numerous suggestions for using these 
    systems for sensing applications. However, there is an open debate as to whether or not this 
    sensitivity advantage is negated by additional noise in the system. 
    We show that an EP sensor's imprecision in measuring a generalized force
    is \textit{independent} of its operating point's proximity to the EP. 
    That is because frequency noises of fundamental origin in the sensor --- due to quantum and thermal fluctuations
    --- increase in a manner that exactly cancels the benefit of increased resonant frequency sensitivity near the EP.
    So the benefit of EP sensors is limited to the regime where
    sensing is limited by technical noises.
    Finally, we outline an EP sensor with phase-sensitive gain that does have an advantage even if limited
    by fundamental noises. 
\end{abstract}

\maketitle

\textit{Introduction.} A classical exceptional point (EP) system with modes $a$ and $b$ is governed by
\cite{Miri19}
\begin{equation}\label{eq:generalEpOde}
    \begin{bmatrix} \dot{a} \\ \dot{b}\end{bmatrix} 
    = -i \mathbf{H}_\text{EP} \begin{bmatrix} a \\ b \end{bmatrix}
\end{equation}
where by definition, the matrix $\mathbf{H}_\text{EP}$ has a degenerate spectrum with a single complex 
eigenvalue $\Omega_\text{EP} - i \gamma_\text{EP}$. Any physical realization of such a system
has to be open, i.e. featuring gain and loss.

A classical EP system is said to be an EP sensor if the degeneracy of $\vb{H}_\t{EP}$ is lifted by a perturbation
of the form $\mathbf{H}_\text{EP} + \epsilon \mathbf{H}_\text{pert}$, such that for
$\epsilon \ll 1$, the eigen-frequencies bifurcate as $\sqrt{\epsilon}$;
$\epsilon = 0$ is termed the exceptional point.
Their claimed utility as a sensor is that if $\epsilon$ is proportional to some external parameter of interest, then,
in contrast to conventional sensors whose eigenfrequencies bifurcate as $\epsilon^1$, EP sensors exhibit enhanced
sensitivity because of the $\epsilon^{1/2}$ scaling near the EP 
\cite{Wang22,Chen17,Liu16,Rosa21,Sakhdari22,Hajizadegan19,Farhat20,Tang23,Li23}.

However, the efficacy of a sensor is not decided by large sensitivity to the quantity
being sensed, but rather by its imprecision, which 
depends on both its sensitivity and added noise. Since the sensitivity of EP sensors is
well understood, controversy has swirled around the fundamental noise inherent
in EP sensors. Some say these sensors offer a fundamental advantage \cite{Zhang19,Zhong19,Kononchuk22}, 
while others disagree \cite{Langbein18,Lau18,Chen19,Wang20,Duggan22,Ding23}. 

Part of the controversy is due to the restricted validity of EP sensing models. If an EP sensor operates near an EP with less gain than loss, it has no macroscopic mode amplitudes unless it is excited externally, and can be analyzed as a parameter estimation problem as in refs. \cite{Lau18,Chen19,Ding23}. However, when an EP sensor is operated near an EP with equal gain and loss, it acts as a two-mode laser above threshold and must be treated differently.

In fact, in an experiment demonstrating an EP sensor above its lasing threshold \cite{Wang20} --- 
a Brillouin ring laser gyroscope --- excess noise was found to exactly cancel any enhancement from the gyroscope's frequency splitting near its EP. 
Our analysis shows that this behavior is characteristic of all EP sensors operated above threshold.

\textit{Methods of EP Sensing.} We break down the class of all EP sensors into sub-categories of sensors, which are more amenable to individual treatment. 
Following existing convention, we assume that these sensors rely on phase-insensitive gain and have reciprocal coupling, i.e. the coupling rate from $a$ to $b$ is the same as from $b$ to $a$. This assumption will be relieved later.

EP sensors are primarily divided into passive EP sensors --- those with no gain --- and active EP sensors --- 
which have gain. 
Refs. \cite{Langbein18,Chen19,Ding23} all conclude that passive EP sensors have no observable $\sqrt{\epsilon}$ bifurcation and thus no fundamental sensing improvement over traditional schemes. 
(We verify these results using the quantum noise formalism of this paper in \cref{subsec:passiveEpSys}.)

Since passive EP sensors have no fundamental sensing benefit, we consider active EP sensing schemes. 
The classical equations of motion for an active EP system are
\begin{equation}\label{eq:activeEp}
\begin{split}
    \dot{a} &= -i \Omega_0 a - \gamma a + \frac{1}{2}(g + \gamma) b \\
    \dot{b} &= -i \Omega_0 b + g b - \frac{1}{2}(g + \gamma) a
\end{split}
\end{equation}
and this system has one eigenfrequency, with real part $\Omega_\text{EP} = \Omega_0$ and imaginary part $\gamma_\text{EP} = (\gamma - g)/2$.
These systems undergo a lasing transition around $\gamma_\t{EP} = 0$ (``loss'' = ``gain''): 
marginally above threshold ($\gamma_\t{EP} = 0^-$), or beyond it ($\gamma_\t{EP} < 0$), the mode amplitudes have 
a finite value. 
Below threshold ($\gamma_\t{EP} > 0$), there is no finite mode amplitude and
the sensor is operated by an external drive. As shown in ref. \cite{Chen19} 
(see also \cref{sec:generalEpSystems}), below threshold active EP sensors have no 
fundamental sensing advantage.

The only remaining cases to be analysed are active EP sensors above threshold with balanced gain and loss, 
or beyond threshold with more gain than loss\footnote{One could also consider an EP sensor operated ``at threshold'' with balanced gain and loss but no macroscopic field. However, such a system is unstable and noise will eventually cause it to acquire a macroscopic field, reverting it to the ``above threshold'' case.}. 
EP sensors operating beyond threshold are unstable as one of their modes is continually amplified. 
However, in any physical system, saturation or other nonlinear mechanisms conspire to eventually stabilize
this runaway; i.e. the gain $g$ is decreased, or the loss $\gamma$ increased, until the system reverts to a 
stable state above threshold with balanced gain and loss rates, i.e. $g = \gamma$ and $\gamma_\text{EP} = 0$.

Such above threshold EP systems, i.e. EP systems with $\gamma_\t{EP}=0$, are ``PT-symmetric''.
Their utility as a sensor is qualitatively different depending on which elements of the coupling matrix 
are perturbed; the diagonal elements, corresponding to the modes' frequencies and gain or loss rates, 
or the off-diagonal elements corresponding to the coupling between the modes. 
\Cref{sec:perts} analyzes perturbations to the diagonal elements. These perturbations either fail to lift the degeneracy, lead to a linear-in-$\epsilon$ response, 
push the system below threshold where the analysis of \cite{Chen19} and \cref{subsec:activeEps} applies, or are equivalent to perturbing the off-diagonal 
matrix elements.

In sum, the only possible type of EP sensor that has not been ruled out by previous analyses and has an enhanced signal near the EP is a PT-symmetric EP sensor where the quantity being sensed alters the coupling between the sensor's modes. 
The classical mode amplitudes of such a sensor satisfy
\begin{equation}\label{eq:epOde}
\begin{split}
    \dot{a} &= -i \Omega_0 a - \gamma a + \gamma (1 + \epsilon) b \\
    \dot{b} &= -i \Omega_0 b + \gamma b - \gamma (1 + \epsilon) a
\end{split}
\end{equation}
where $\epsilon$ is the perturbation being detected by the sensor relative to it's EP with $\Omega_\text{EP} = \Omega_0$ and $\gamma_\text{EP} = 0$. We must have $\epsilon >0$ for the sensor to be stable, otherwise one of the system's normal modes is continually amplified, which is unphysical.
So the sensor emits an output mode with a finite amplitude whose frequency depends on $\epsilon$, which 
is, by assumption, coupled to the sensed quantity. 
Thus, it is the frequency of the output mode that is expected to feature the $\epsilon^{1/2}$ scaling that leads
to a sensing advantage. In order to verify whether this sensitivity advantage holds, it is important to 
analyse how fundamental frequency noise in such a system scales with $\epsilon$. In the sequel we 
show that in fact quantum (and thermal) frequency noise also scales as $\epsilon^{1/2}$, nullifying any
reduction of PT-symmetric EP sensors' imprecision in measuring $\epsilon$.

\begin{figure}[t!]
    \centering
    \includegraphics[width=0.9\columnwidth]{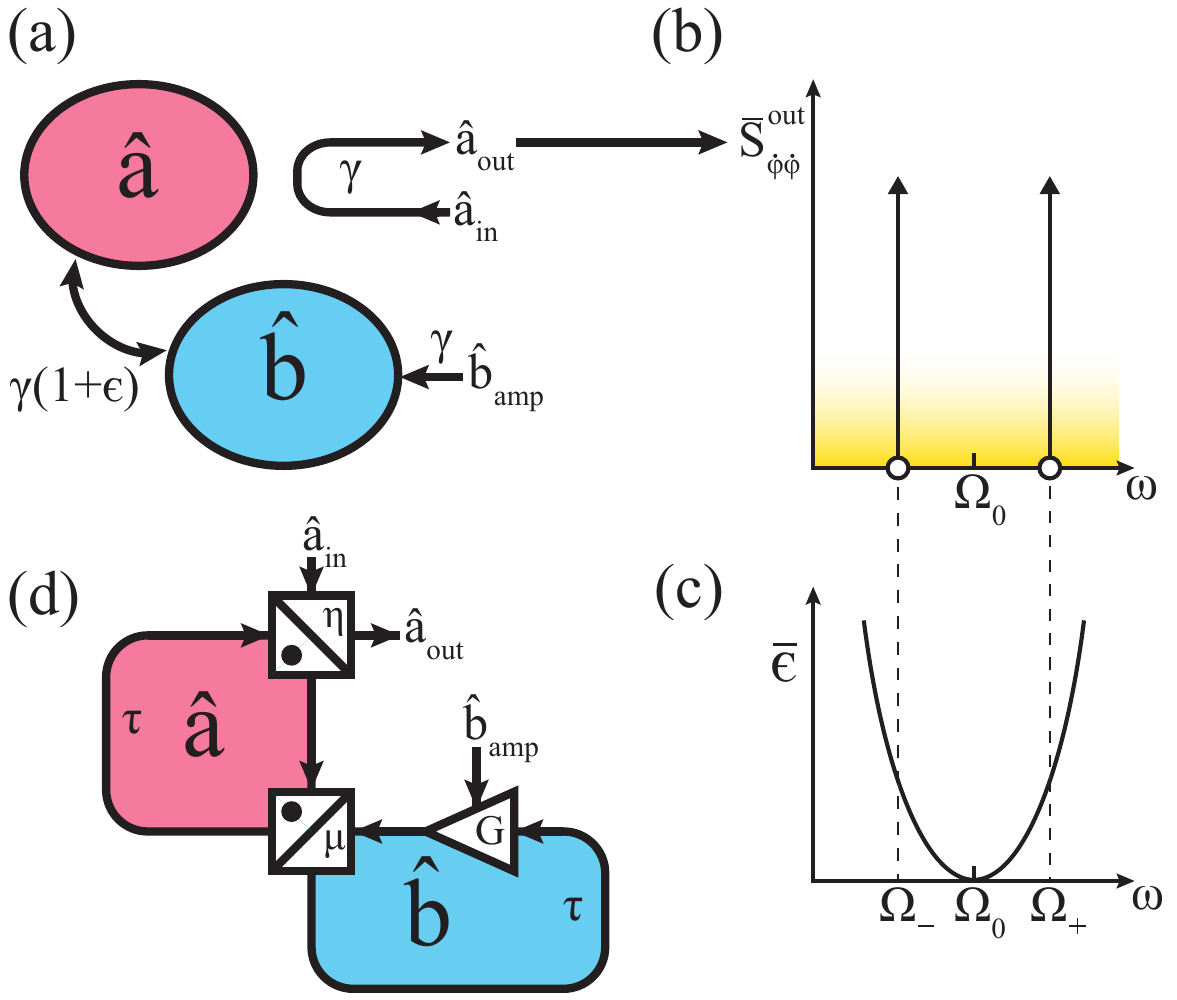}
    \caption{\label{fig:epSensorSchematic} A schematic of PT-symmetric EP sensors used to detect a weak signal. Mode $\hat{a}$ is coupled to the output by a decay rate $\gamma$, while mode $\hat{b}$ is amplified at a rate $\gamma$, which introduces a noise mode $\hat{b}_\t{amp}$. 
    The modes are coupled to each other at the rate $\gamma(1+\epsilon)$, where $\epsilon$ is proportional to the weak signal being measured. 
    (a) shows a Markovian EP sensor with dynamics described by \cref{eq:epQuantumOde}. 
    (b) is the output frequency spectrum, a pair of delta-function spikes surrounded by white noise. The locations of the delta functions in (b) are determined by the small quantity being sensed and split quadratically, as in (c). 
    (d) shows a physical implementation of the sensor using a pair of beam splitters (or partially transmissive mirrors), two delay lines, and an amplifier. The weak signal alters the transmissivity of the beam splitter coupling modes $\hat{a}$ and $\hat{b}$.}
\end{figure}

\emph{Fundamental frequency noise in PT-symmetric EP sensors.} 
Since EP sensors will ultimately be limited by quantum (and thermal) fluctuations, 
the classical model of \cref{eq:epOde} is incomplete as far as noise performance is concerned.
A minimal quantum model that reproduces \cref{eq:epOde} in an expectation value sense can be obtained
by promoting the mode amplitudes to operators and adding appropriate noise terms which ensure
the preservation of appropriate commutation relations between the 
operators \cite{Loui61,GorWalk63b,Stre82,Collett84,Gardiner85}.
We also decompose $\epsilon$ into its mean value and fluctuations as $\epsilon = \bar{\epsilon} + \delta \epsilon$,
where $\bar{\epsilon} \equiv \avg*{\epsilon}$.
This decomposition will allow us to consider the EP sensor as a probe for weak forces (modeled by $\delta \epsilon$) or 
as a probe to estimate the unknown parameter $\bar{\epsilon}$ .
The resulting equations of motion for the quantized mode amplitudes are
\begin{equation}\label{eq:epQuantumOde}
\begin{split}
    \dot{\hat{a}} &= (-i \Omega_0 - \gamma) \hat{a} + \gamma (1 + \bar{\epsilon} + \delta \epsilon) \hat{b} 
        + \sqrt{2 \gamma}\, \delta\hat{a}_\text{in} \\
    \dot{\hat{b}} &= (-i \Omega_0 + \gamma) \hat{b} - \gamma (1 + \bar{\epsilon} + \delta \epsilon) \hat{a} 
        + \sqrt{2 \gamma}\, \delta\hat{b}^\dagger_\text{amp}
\end{split}
\end{equation}
where the input modes $\hat{a}_\text{in}$ and $\hat{b}_\text{amp}$ have zero expectation value
and are purely noisy.
Their noise properties are quantified by the symmetrized, double-sided spectra 
$\bar{S}_{qq}^\text{in} = \bar{S}_{pp}^\text{in} = \tfrac{1}{2} + n_\text{in}$ and $\bar{S}_{qq}^\text{amp} = \bar{S}_{pp}^\text{amp} = \tfrac{1}{2} + n_\text{amp}$, 
for their amplitude and phase quadratures, $\hat{q}_{\text{in},\text{amp}} \equiv (\hat{a}_{\text{in},\text{amp}}^\dagger + \hat{a}_{\text{in},\text{amp}})/\sqrt{2}$ and $\hat{p}_{\text{in},\text{amp}} \equiv i(\hat{a}_{\text{in},\text{amp}}^\dagger - \hat{a}_{\text{in},\text{amp}} )/\sqrt{2}$ respectively \cite{Clerk10}.  
In these spectra, $n_\t{in,amp}$ are the average thermal occupation describing thermal fluctuations entering through 
these modes, which can be zero; however quantum (vacuum) noise gives rise to the $\tfrac{1}{2}$ terms, which is
unavoidable in any consistent quantum description of a PT-symmetric EP sensor.

In order to observe the modes' frequency shifts, it is necessary to consider the output mode emitted by
the system.
The natural candidate is the output mode \cite{Gardiner85} $\hat{a}_\t{out} = \sqrt{2 \gamma} \hat{a} 
- \hat{a}_\t{in}$, leaking out through the lossy element, as depicted in \cref{fig:epSensorSchematic}(a).
(A similar input-output relation exists for the mode $\hat{b}_\t{out}$, but this is the amplifier's out-coupled 
mode and is usually unobservable.) In a physical EP sensor, the sensor's loss comes from a beam-splitter
interaction, and the output mode is transmitted out of the sensor's feedback loop through this beam-splitter,
as in \cref{fig:epSensorSchematic}(d).

Taking quantum expectation values of \cref{eq:epQuantumOde} gives the classical amplitudes 
$a \equiv \avg{\hat{a}}$ and $b \equiv \avg*{\hat{b}}$
\begin{equation}\label{eq:eomSols}
\begin{split}
    a(t) &= a_+ e^{-i \Omega_+ t} + a_- e^{-i \Omega_- t} \\
    b(t) &= b_+ e^{-i \Omega_+ t} + b_- e^{-i \Omega_- t},
\end{split}
\end{equation}
oscillating at the normal-mode frequencies $\Omega_\pm = \Omega_0 \pm \gamma \sqrt{\bar{\epsilon} (2 + \bar{\epsilon})}$
featuring the $\bar{\epsilon}^{1/2}$ scaling around the EP. This is depicted in \cref{fig:epSensorSchematic}(c).
Here, $a_\pm$ and $b_\pm$ are complex-constants.
As discussed in detail later in the paper, the EP sensor has quadrature spectra with poles at $\Omega_\pm$, which will build up coherent oscillations from noise, as in a laser. 
The coefficients $a_\pm$ and $b_\pm$ will then be determined by saturation effects, not by initial conditions, and will be independent of $\epsilon$ as long as the saturation mechanism is.

The noise in the system is characterized by the operators $\delta \hat{a} \equiv \hat{a} - \avg{\hat{a}}$, 
$\delta \hat{b} \equiv \hat{b} - \avg*{\hat{b}}$. Their equations of motion follow from linearizing 
\cref{eq:epQuantumOde}:
\begin{equation}\label{eq:epCoupledMode}
\begin{split}
    \delta \dot{\hat{a}} &= -i \Omega_0 \delta\hat{a} - \gamma \delta\hat{a} + \gamma (1 + \bar{\epsilon}) \delta\hat{b} + \gamma b\, \delta \epsilon
        + \sqrt{2 \gamma} \, \delta \hat{a}_\text{in} \\
    \delta \dot{\hat{b}} &= -i \Omega_0 \delta\hat{b} + \gamma \delta\hat{b} - \gamma (1 + \bar{\epsilon}) \delta\hat{a} - \gamma a\, \delta \epsilon
        + \sqrt{2 \gamma} \, \delta \hat{b}^\dagger_\text{amp}.
\end{split}
\end{equation}
These fluctuations in the internal modes leak out according to 
$\delta \hat{a}_\text{out} = \sqrt{2 \gamma} \, \delta \hat{a} - \delta\hat{a}_\text{in}$. 
To determine the system's output fluctuations, we solve \cref{eq:epCoupledMode} in the 
frequency domain, assuming that the fluctuation dynamics are much faster than the mean dynamics of \cref{eq:epOde}. 
To simplify the equations, we assume that we operate the sensor near resonance and near the EP
such that $\omega - \Omega_0 \ll \gamma$ and $\bar{\epsilon} \ll 1$. 
(We show in \cref{sec:noWeakStationary} that removing these assumptions does not alter the results presented here.)
The sensor's output phase quadrature fluctuations are given by
\begin{equation}\label{eq:aOut}
\begin{split}
    \delta \hat{p}_\text{out}[\omega] = \frac{2 \gamma^2}{(\omega - \Omega_-)(\omega - \Omega_+)} \Big[ \delta \hat{p}_\text{in}[\omega] + \delta \hat{p}_\text{amp}^\dagger[\omega] \\
    \quad + \sqrt{2\gamma} \Big( p_a^- \delta \epsilon[\omega - \Omega_-] + p_a^+ \delta \epsilon[\omega - \Omega_+] \Big) \Big],
\end{split}
\end{equation}
where $p_a^\pm$ are the phase quadratues of the constants $a_\pm$. 
A similar result holds for the amplitude quadrature.
Given the EP sensor's output fluctuations, we can now evaluate its performance.

\textit{Parameter Estimation.} We first consider using the EP sensor to estimate the unknown parameter $\bar{\epsilon}$. In this case, we assume $\epsilon$ is varying slowly, and we can ignore the $\delta \epsilon$ terms in \cref{eq:aOut}.
Considering the EP sensor's spectrum near a resonant frequency and near the EP, we define $\Delta \omega_\pm \equiv \omega - \Omega_\pm$ and invoke the near-resonant approximation to work to leading order in $\Delta \omega_\pm/\gamma$. We continue to assume that we are near enough to the EP that we can work to leading order in $\bar{\epsilon}$. We need to be careful about how these approximations interact in the denominator of \cref{eq:aOut}: since PT-symmetric EP sensor will operate around some small constant value of $\bar{\epsilon}$, but a frequency measurement will only be affected by frequency noise infinitesimally close to resonance, $\Delta \omega_\pm$ vanishes faster than $\bar{\epsilon}$.
With these approximations, the output phase quadrature spectrum takes the form
\begin{equation}\label{eq:SppApprox}
    \bar{S}_{pp}^\text{out} [\Omega_\pm + \Delta \omega_\pm] = \frac{\gamma^2(1 + 2 n_\text{th}) } 
    {2 \bar{\epsilon} \Delta \omega_\pm^2}.
\end{equation}
In this expression, quantum zero-point fluctuations contribute the constant term in the numerator, while thermal noise 
in the amplifier and in-coupled mode contribute identically via the average thermal occupation $n_\t{th} \equiv (n_\text{amp} + n_\text{in})/2$. 

Since it is the frequency shift $\Omega_\pm (\bar{\epsilon})$ that is sensed, it is the frequency noise corresponding
to the above spectrum that is relevant. The sensor's output frequency spectrum is given by
\begin{align}
    \bar{S}_{\dot{\varphi} \dot{\varphi}}[\Omega_\pm + \Delta\omega_\pm] 
    = \frac{\Delta \omega_\pm^2}{4 \gamma |a_\pm|^2} \bar{S}_{pp}^\text{out}[\Omega_\pm + \Delta\omega_\pm] 
\end{align}
Converting \cref{eq:SppApprox} into an equivalent frequency spectrum, we find
\begin{equation}\label{eq:stEp}
    \bar{S}_{\dot{\varphi} \dot{\varphi}}[\Omega_\pm + \Delta\omega_\pm] = 
    \frac{\gamma (1 + 2 \bar{n}_\text{th})}{8 |a_\pm|^2 \bar{\epsilon}}.
\end{equation}
This can be recognized as the Schawlow-Townes spectrum \cite{LouSud23}, 
with the $1/\bar{\epsilon}$ factor interpreted as a Petermann factor, which generically describes excess laser frequency noise in multimode lasers with non-orthogonal modes \cite{Siegman00,Wang20}.

\begin{figure}[t!]
    \centering
    \includegraphics[width=0.95\columnwidth]{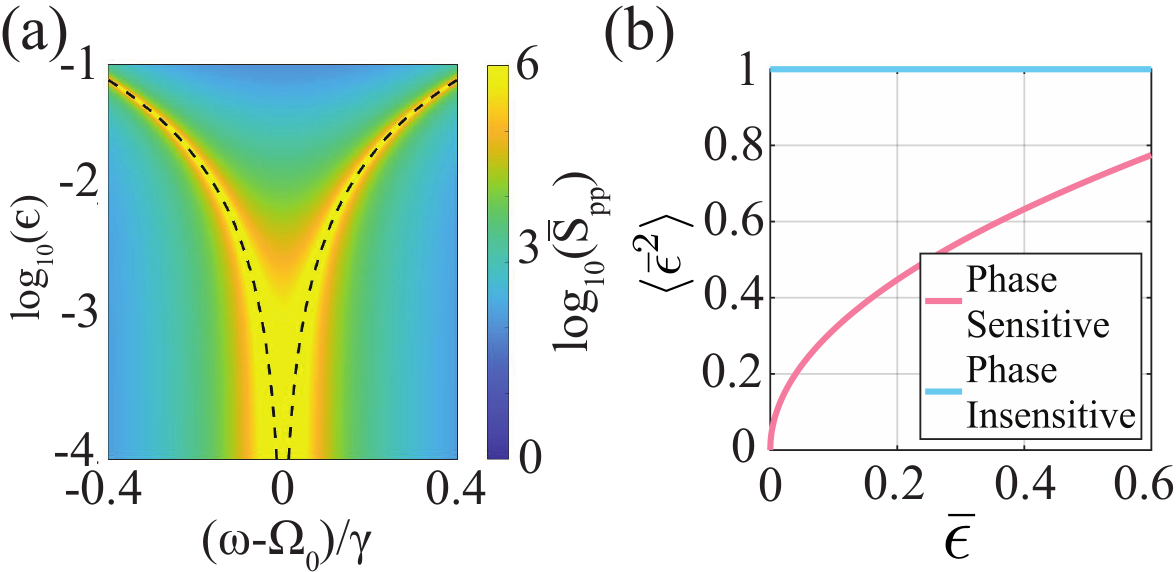}
    \caption{\label{fig:epSpectrum} Markovian PT-symmetric EP sensors' output phase quadrature spectrum, $\bar{S}_{pp}^\t{out}[\omega]$. 
    (a) The phase quadrature spectrum of a sensor with a phase-insensitive gain as a function of frequency and fractional distance to the EP, $\epsilon$. 
    The dashed black line shows the sensor's resonant frequencies scaling as $\sqrt{\epsilon}$, giving rise to increased sensitivity for $\epsilon \ll 1$. 
    But the fundamental noises (quantum and thermal) also increases with the same scaling so that resonance locations are no easier to 
    distinguish than when the sensor is farther from the EP. 
    (b) While the ability of an EP sensor to estimate the value of small parameter $\bar{\epsilon}$ is independent of $\bar{\epsilon}$ for a sensor with a phase-insensitive gain, sensors with phase-sensitive gain are more sensitive as $\bar{\epsilon}$ decreases. In these plots, we take $\bar{n}_\t{th} = 0$. 
    In (b), we take $\Delta\omega_\text{meas}/(16 \pi \gamma_a |a_\pm|^2) = 1$ and $\gamma_b \Delta\omega_\text{meas}^2/(6 \gamma_a^2 (\gamma_a+\gamma_b)) = 1$.}
\end{figure}

Importantly, the frequency noise given by \cref{eq:stEp} increases as we approach the EP
(i.e. $\bar{\epsilon} \rightarrow 0$), meaning that PT-symmetric EP sensors pay a penalty in the form 
of increased quantum and thermal noise near the EP, as shown in \cref{fig:epSpectrum}a. 

The scaling of the frequency noise $\sqrt{\bar{S}_{\dot\phi \dot \phi}} \sim 1/\sqrt{\bar{\epsilon}}$ 
in fact \emph{precisely nullifies} the $\sqrt{\bar{\epsilon}}$ scaling in the frequency sensitivity. 
To wit, consider the scenario where the unknown parameter is to be estimated from, say the difference in
frequency of the modes (measuring a single frequency does not change the conclusion): then the 
sensitivity to $\bar{\epsilon}$ is
$\mathcal{S} = |\partial (\Omega_+ - \Omega_-)/\partial \bar{\epsilon}|$, whereas the noise 
is $\mathcal{N} = \sqrt{(\bar{S}_{\dot{\varphi} \dot{\varphi}}[\Omega_+ + \Delta\omega] + \bar{S}_{\dot{\varphi} \dot{\varphi}}[\Omega_- + \Delta\omega])\Delta \omega_\text{meas}/(2 \pi)}$ (here $\Delta \omega_\t{meas}$ is the measurement bandwidth). 
The imprecision in the estimate of $\bar{\epsilon}$ is \cite{Braunstein96}, $\avg*{\bar{\epsilon}_\t{imp}^2}
\equiv \mathcal{N}/\mathcal{S}$, which in this case
\begin{equation}\label{eq:ptSnr}
    \langle \bar{\epsilon}^2_\text{imp} \rangle
    =  \sqrt{\frac{(1 + 2 \bar{n}_\text{th}) \Delta \omega_\text{meas}}{16 \pi \gamma |a_\pm|^2}}
\end{equation}
is \textit{independent} of the sensor's proximity to the EP. 
Thus, operating a sensor near a PT symmetric EP offers no fundamental imprecision reduction for parameter estimation
since the sensitivity and noise are enhanced equally.

\textit{Weak Force Sensing.} Instead of using the EP sensor to estimate the value of $\bar{\epsilon}$, we could instead consider using it to measure a weak force coupled to $\delta \epsilon$. Assuming the fluctuations in $\delta \epsilon$ are weak stationary near the EP, the sensor's output quadrature spectra can be computed from \cref{eq:aOut}
using the known spectra of the fundamental input noises. Near the EP, near resonance the output quadrature spectra are
given by
\begin{equation}\label{eq:weakForceMeas}
    \bar{S}_{xx}^\text{out}[\omega] = 4 \gamma^4\frac{(1 + 2n_\text{th}) + 2\gamma(|x_-|^2 + |x_+|^2) \bar{S}_{\epsilon \epsilon}[\omega - \Omega_0]}{(\omega - \Omega_-)^2(\omega - \Omega_+)^2}.
\end{equation}
Here $x \in \{q,p\}$ is the amplitude or phase quadrature.
From \cref{eq:weakForceMeas}, we see that as in the parameter estimation case, there is no sensing advantage to using an EP sensor to measure a weak force: the fluctuations in the weak force $\bar{S}_{\epsilon \epsilon}$ and the fundamental
noises $\sim (1+2n_\t{th})$, are both transduced to the output via the same response $1/(\omega-\Omega_\pm)^2$, so
that any scaling of the response with $\epsilon$ is immaterial in distinguishing the weak force from 
the fundamental noises. Relaxing the assumptions of weak stationarity and/or proximity to resonance and/or EP 
do not alter this conclusion (see \cref{sec:noWeakStationary}). 

In sum, neither parameter estimation nor weak-force-sensing benefits from using an EP sensor 
because unavoidable quantum frequency noises in an EP system precisely nullify the enhanced sensitivity.

\textit{Practical Advantage of PT-Symmetric Sensors.} While PT-symmetric sensors do not offer any advantage 
in overcoming the limitation set by fundamental noises of quantum and thermal origin, they can potentially 
be advantageous when the sensor is limited by technical noises. 
By technical noise, we mean any noise that is uncorrelated with the fundamental noises and independent 
of $\epsilon$ (an example is apparent frequency noise in the readout of the sensor). Suppose technical noise
is characterized by its spectrum $\bar{S}_{\dot{\varphi} \dot{\varphi}}^\t{tech}$, the imprecision with which 
$\bar{\epsilon}$ can be estimated is
\begin{equation}
    \langle \bar{\epsilon}^2_\t{imp} \rangle = \sqrt{\frac{\bar{S}_{\dot{\varphi} \dot{\varphi}} + \bar{S}_{\dot{\varphi} \dot{\varphi}}^\text{tech}}
    {2 \gamma^2 /\bar{\epsilon}}}.
\end{equation}
If technical noises dominate (i.e. $\bar{S}_{\dot{\varphi} \dot{\varphi}}^\text{tech} 
\gg \bar{S}_{\dot{\varphi} \dot{\varphi}}$), it is advantageous to operate the sensor by approaching the EP;
however since the spectrum of fundamental noises is characrerized by 
$\bar{S}_{\dot{\varphi}\dot{\varphi}} \sim 1/\bar{\epsilon}$, approaching the EP will eventualy make the 
fundamental noises dominate, saturating the imprecision in estimating $\bar{\epsilon}$ to the value given in \cref{eq:ptSnr}.

\textit{Effects of non-Markovian dynamics.} The analysis above considers PT-symmetric EP sensors in the Markovian limit where the coupled-mode equations of \cref{eq:epQuantumOde} are valid. This case is depicted in \cref{fig:epSensorSchematic}(a). While the Markovian assumption leading to these input-output equations \cite{Collett84} is often an excellent assumption, physical optical systems are more accurately described by a set of non-Markovian coupled mode equations. Such a physical system is depicted in \cref{fig:epSensorSchematic}(d).

In \cref{sec:nonMarkov}, we consider the exact dynamics of a non-Markovian PT-symmetric sensor 
with finite propagation delays and find that such a sensor's frequency noise near resonance, 
relevant for estimating $\bar{\epsilon}$, is given by
\begin{equation}\label{eq:freqSpecNonMarkov}
    \bar{S}_{\dot{\varphi} \dot{\varphi}} [\Omega_\pm + \Delta\omega_\pm] = \frac{\eta^2 (1 + 2 \bar{n}_\text{th})}{4 (1 - \eta)(2 - \eta)^2 \tau^2 |\alpha_\pm|^2 \bar{\epsilon}}
\end{equation}
where $\eta$ is the transmissivity of the sensor's input-output coupler, and $\tau$ is the round trip time delay for one oscillation for modes $\hat{a}$ and $\hat{b}$. 
The only difference compared to \cref{eq:stEp} is that the Markovian decay rate $\gamma$ is replaced by 
$\eta/(\sqrt{1-\eta}(2-\eta)\tau)$; indeed, in the Markovian limit $\eta,\tau \rightarrow 0$ 
with $\eta/(2\tau) = \gamma$ held fixed, \cref{eq:stEp,eq:freqSpecNonMarkov} are equivalent. 
Importantly, the $1/\sqrt{\bar{\epsilon}}$ scaling in the noise is again present; so
non-Markovan dynamics still respects the conclusion that PT-symmetric EP 
sensors do not offer a fundamental advantage.

\textit{Quantum-enhanced EP sensors.} Since quantum noise is found to be responsible for nullifying any advantage
arising from proximity to the EP, it is conceivable that by engineering the quantum states of the sensor, an 
advantage can be recovered. A strategy to limit excess noise is to use  
phase-sensitive gain in the PT-symmetric sensor instead of a phase-insensitive one. 
A phase-sensitive PT-symmetric sensor is described by the coupled-mode equations [\cref{sec:phSensMarkovDynamics}]
\begin{equation}\label{eq:phSensLangevain}
\begin{split}
    \dot{\hat{a}} =& -i \Omega_0 \hat{a} - \gamma_a \hat{a} + \gamma_a (1 + \bar{\epsilon} + \delta \epsilon) \hat{b} + \sqrt{2 \gamma_a} \delta \hat{a}_\text{in}\\
    \dot {\hat{b}} =& -i \Omega_0 \hat{b} + (\gamma_a-r) \hat{b} + e^{-2i\Omega_0 t}r \hat{b}^\dagger -\gamma_a (1+\bar{\epsilon} + \delta \epsilon) \hat{a} \\
    & + \sqrt{2 (\gamma_a + \gamma_b - r)}\, \delta \hat{b}_\text{amp}^\dagger + \sqrt{2 \gamma_b} \, \delta \hat{b}_\text{in}.
\end{split}
\end{equation}
We assume that the out-coupled mode $\hat{a}_\text{out}$ is observable, whereas $\hat{b}_\text{out}$ 
is not.
Here, $r$ is the rate of phase-sensitive amplification; for stability, $0 \leq r \leq \gamma_a + \gamma_b$.

The most optimistic scenario is one where added noise is least, which is the case where the gain is purely
phase-sensitive, i.e. $r = \gamma_a+\gamma_b$. 
(\Cref{sec:phaseSensitveSensor} contains a more general analysis.)
The output phase spectrum near resonance is given to leading order in $\Delta \omega_\pm$ and $\bar{\epsilon}$ by 
\begin{equation}\label{eq:SppOutPhSens}
    \bar{S}_{pp}^\text{out} [\Omega_\pm + \Delta \omega_\pm] = \frac{\gamma_b}{2(\gamma_a+\gamma_b)} (1 + 2 \bar{n}_\text{th}).
\end{equation}
Since the system's output frequency spectrum is proportional to its output phase quadrature spectrum, the frequency spectrum will also be independent of the proximity to the PT-symmetric EP and this sensor will have an imprecision that decreases as $\sqrt{\bar{\epsilon}}$. 
The right panel of \cref{fig:epSpectrum} shows the phase quadrature spectrum of a phase-sensitive EP sensor. 

We could also consider using the phase-sensitive EP sensor for weak force sensing. We show in \cref{sec:phSensWeakForce} that there is no sensing enhancement in this regime, as in the phase-insensitive case. 
Note that the advantage due to this phase-sensitive strategy is different from strategies relying on
non-reciprocal coupling of the modes \cite{Lau18}. 

\textit{Conclusion.} 
PT-symmetric EP sensors, just like other physical systems, are fundamentally affected by quantum noise. Accounting for
the quantum noise in a self-consistent fashion, and plugging a range of potential loopholes, we show that PT-symmetric
EP sensors do not offer any advantage for parameter estimation or weak-force sensing if the sensor is limited
by fundamental (i.e. quantum and thermal) noises. That is because these noises scale in exactly the opposite
manner as the scaling of the sensitivity near the EP, thus nullifying any improvement in signal-to-noise ratio.
So the advantage of such sensors is confined to the scenario where they are limited by technical noises.
Beating the limits set by quantum noise requires quantum resources: we outline a phase-sensitive generalization
of an EP sensor that does confer an advantage by harnessing the square-root bifurcation near an EP.

\clearpage
\onecolumngrid
\appendix
\tableofcontents

\section{Mathematical Conventions}\label{app:mathDetails}

We adopt the same mathematical conventions as ref. \cite{LouSud23}, detailed in the first appendix of that paper. Reference \cite{LouSud23} did not emphasize the spectra of thermal states, which we briefly discuss here. A thermal state has quadrature spectra given by \cite{Clerk10}
\begin{equation}
    \bar{S}_{qq}[\omega] = \bar{S}_{pp}[\omega] = \frac{1}{2} + n_\text{th}[\omega],
\end{equation}
where $\bar{n}_\t{th}[\omega]$ is the Bose-Einstein occupation number for a thermal state at frequency $\omega$ given by
\begin{equation}
    n_\text{th}[\omega] = \frac{1}{\exp(\frac{\hbar \omega}{k_\text{B} T})-1},
\end{equation}
where $\hbar$ is the reduced Planck constant, $k_\t{B}$ is Boltzmann's constant, and $T$ is the system's temperature.
In this paper, we assume that the thermal occupation number, $n_\t{th}[\omega]$, does not vary appreciably across the frequency range of interest, and we treat it as a constant with a value $n_\t{th}[\omega] = n_\t{th}$.

\section{General Exceptional Point Systems}\label{sec:generalEpSystems}

We consider the properties of two-mode exceptional point sensors, which are not necessarily PT-symmetric. We divide these sensors into several categories depending on the loss and amplification of each mode.

\begin{enumerate}
    \item Both modes are amplified. This is unphysical as it will cause the modes' amplitudes to diverge, so we do not consider this case further. It also does not provide a mechanism to extract a signal out of the sensor.
    \item Both modes are lossy.
    \item One mode is lossy and the other is amplified. 
\end{enumerate}
It is also possible for one or both modes to have both gain and loss, but this will only add noise to the system relative to the case where each mode is purely amplified or lossy with the same net amplification or loss rates. We now consider the properties of EP sensors in the later two cases above. We will find that when both modes are lossy, the system does not have an observable square-root dependence on small perturbations. Similarly, in the case where one mode is lossy while the other is amplified, the system has an observable square-root dependence on small perturbations only if the amplification and loss rates are equal, i.e. if the sensor is PT-symmetric. Thus, we conclude that the only type of EP sensor with an observable square-root dependence on small perturbations is a PT-symmetric EP sensor.

\subsection{Passive Exceptional Point Systems} \label{subsec:passiveEpSys}

An exceptional point sensor with no amplification is described by
\begin{equation}\label{app:passiveEpEom}
\begin{split}
    \partial_{t} \hat{a} &= - i \Omega_0 \hat{a} - \gamma_a \hat{a} + \frac{1}{2}|\gamma_a - \gamma_b| (1 + \epsilon) \hat{b} + \sqrt{2 \gamma_a} \, \hat{a}_\text{in} \\
    \partial_{t} \hat{b} &= - i \Omega_0 \hat{b} - \gamma_b \hat{b} - \frac{1}{2}|\gamma_a - \gamma_b| (1 + \epsilon) \hat{a} + \sqrt{2 \gamma_b} \, \hat{b}_\text{in}
\end{split}
\end{equation}
Here, $\gamma_a$ ($\gamma_b$) is the decay rate of mode $\hat{a}$ ($\hat{b}$). The mode $\hat{b}_\text{in}$ is a noise from the loss associated with mode $\hat{b}$. This system's normal mode frequencies are given by
\begin{equation}
    \Omega_\pm^\text{loss} = \Omega_0 - \frac{i}{2} (\gamma_a + \gamma_b) \pm \frac{1}{2}|\gamma_a - \gamma_b| \sqrt{\epsilon (2 + \epsilon)},
\end{equation}
which have the desired square-root splitting with $\epsilon$ near the EP at $\epsilon = 0$. While we consider the system to be perturbed about the EP by perturbing its coupling, the perturbation could instead be to the system's resonant frequencies or decay rates. These other perturbations do not alter the fundamental results here, so we consider only coupling perturbations for simplicity.

Using the boundary condition $\hat{a}_\t{out} = \sqrt{2 \gamma}\, \hat{a} - \hat{a}_\t{in}$ and Fourier transforming, we find that the sensor's output mode $\hat{a}_\t{out}$ is given by
\begin{equation}\label{app:passiveEpTfs}
    \hat{a}_\text{out}[\omega] = - \hat{a}_\text{in}[\omega] - \frac{2\gamma_a (\gamma_b - i (\omega-\Omega_0))}{\frac{1}{4}(\gamma_a + \gamma_b)^2 + (\omega - \Omega_-^\text{loss})(\omega - \Omega_+^\text{loss})}\hat{a}_\text{in}[\omega] + \frac{(1 + \epsilon)\sqrt{\gamma_a \gamma_b} |\gamma_a - \gamma_b|}{\frac{1}{4}(\gamma_a + \gamma_b)^2 + (\omega - \Omega_-^\text{loss})(\omega - \Omega_+^\text{loss})}\hat{b}_\text{in}[\omega].
\end{equation}
We take the transfer functions $H_a$ and $H_b$ to be defined by $\hat{a}_\text{out}[\omega] = H_a[\omega] \hat{a}_\text{in}[\omega] + H_b[\omega] \hat{b}_\text{in}[\omega]$. They can be read off from \cref{app:passiveEpTfs}.

For the passive EP sensor, we can measure the transfer functions $H_a$ and $H_b$ by driving the system with a coherent tone at the $\hat{a}_\t{in}$ or $\hat{b}_\t{in}$ input modes and observing its response in the mode $\hat{a}_\t{out}$. While this system has an EP at $\epsilon = 0$, its transfer functions are non-singular at this point and change linearly with small changes in $\epsilon$. Thus, the passive EP sensor does not have the desired square-root sensitivity to $\epsilon$ near its EP.

In addition to measuring the sensor's transfer functions, we can directly measure its output quadrature spectra. Taking the amplitude and phase quadratures to be $\hat{q} \equiv (\hat{a}^\dagger + \hat{a})/\sqrt{2}$ and $\hat{p} \equiv i(\hat{a}^\dagger - \hat{a})/\sqrt{2}$ respectively, we find that the output quadrature spectra are given by
\begin{equation}\label{app:outputSpectraPassive}
    \bar{S}_{xx}^\text{out}[\omega] = |H_a[\omega]|^2 \bar{S}_{xx}^\text{a}[\omega] + |H_b[\omega]|^2 \bar{S}_{xx}^\text{b}[\omega] + 2 \text{Re}\left[ H_a[\omega] H_b[\omega]^* \bar{S}_{xx}^{ab}[\omega] \right]
\end{equation}
where $\hat{x} \in \{\hat{q},\hat{p}\}$ is the phase or amplitude quadrature, $\bar{S}_{xx}^a$ ($\bar{S}_{xx}^b$) is the $\hat{x}$ quadrature spectrum of mode $\hat{a}_\t{in}$ ($\hat{b}_\t{in}$), and $\bar{S}_{xx}^{ab}$ is the cross-correlation spectrum of the $\hat{x}$ quadrature of the modes $\hat{a}_\t{in}$ and $\hat{b}_\t{in}$. The spectra $\bar{S}_{xx}^{a}$, $\bar{S}_{xx}^{b}$, and $\bar{S}_{xx}^{ab}$ are independent of $\epsilon$ since they depend only on the input modes and not on the sensor's dynamics. The only dependence on $\epsilon$ in \cref{app:outputSpectraPassive} is then encoded in the transfer functions. As discussed above, the transfer functions do not diverge for $\epsilon = 0$ so the output quadrature spectra show no square-root splitting or other divergence with $\epsilon$ near the EP.

\subsection{Active Exceptional Point Systems}\label{subsec:activeEps}

Having seen that passive EP sensors do not show an observable square-root splitting from small perturbations, we now consider active EP sensors where one mode is lossy and the other is amplified. Without loss of generality, we assume that mode $\hat{a}$ is the lossy mode. Such a sensor is governed by the coupled equations
\begin{equation}\label{app:activeEpEom}
\begin{split}
    \partial_{t} \hat{a} &= - i \Omega_0 \hat{a} - \gamma \hat{a} + \frac{1}{2}(\gamma + g) (1 + \epsilon) \hat{b} + \sqrt{2 \gamma} \, \hat{a}_\text{in} \\
    \partial_{t} \hat{b} &= - i \Omega_0 \hat{b} + g \hat{b} - \frac{1}{2}(\gamma + g) (1 + \epsilon) \hat{a} + \sqrt{2 g} \, \hat{b}_\text{amp}^\dagger
\end{split}
\end{equation}
Here, $g$ is the amplification rate of mode $\hat{b}$. For a physical sensor, we must have $\gamma \geq g$, otherwise saturation effects will limit the system's gain or increase its losses until it reaches an equilibrium state with $g = \gamma$. As with the passive EP sensor, we consider that the quantity being measured couples to the sensor by perturbing the coupling between the two modes, but similar results hold for more general coupling schemes such as perturbing the gain or loss rates. 

This EP system with gain and loss has normal mode frequencies given by
\begin{equation}
    \Omega_\pm^\text{gl} = \Omega_0 - \frac{i}{2}(\gamma - g) + \frac{1}{2}(g + \gamma)\sqrt{\epsilon (2 + \epsilon)}
\end{equation}
with the characteristic square-root splitting around $\epsilon = 0$.

Using the same boundary condition as in the passive sensing case, and Fourier transforming, we find that $\hat{a}_\t{out}$ is given by
\begin{equation}\label{app:tfsActive}
    \hat{a}_\text{out}[\omega] = -\hat{a}_\text{in}[\omega] - \frac{2 \gamma (g + i(\omega - \Omega_0))}{\frac{1}{4}(\gamma - g)^2 + (\omega - \Omega_-^\text{gl})(\omega - \Omega_+^\text{gl})}\hat{a}_\text{in}[\omega] + \frac{(1 + \epsilon)\sqrt{g \gamma}(g + \gamma)}{\frac{1}{4}(\gamma - g)^2 + (\omega - \Omega_-^\text{gl})(\omega - \Omega_+^\text{gl})}\hat{b}_\text{amp}^\dagger[\omega]
\end{equation}
From \cref{app:tfsActive}, we can read off the transfer functions from the input and amplifier modes to the sensor's output. For $g < \gamma$, the transfer functions, and thus the output quadrature spectra, are non-singular for $\epsilon = 0$ and do not diverge with small values of $\epsilon$. 

Combining the results of \cref{subsec:passiveEpSys} and \cref{subsec:activeEps}, we see that passive EP systems and active EP systems with larger loss rates than amplification rates do not have observable square-root dependencies on small perturbations despite the fact that their complex eigen-frequencies diverge as the square-root of small perturbations. This analysis agrees with the previous analyses of \cite{Chen19,Langbein18,Ding23}. The only remaining possibility of achieving a parametric sensitivity enhancement then comes from active EP systems with equal amplification and loss rates, i.e. PT-symmetric EP sensors.

\section{Perturbations of Diagonal Coupling Matrix Elements}\label{sec:perts}

In the main text, we considered only perturbations to the coupling between modes in a PT-symmetric EP sensor, the off-diagonal elements in the system's coupled-mode matrix. Here, we consider perturbations to the diagonal elements. We consider differential and common-mode perturbations to the system's uncoupled resonant frequencies and gain and loss rates. We will see that perturbations to these diagonal matrix elements either do not lead to the desired $\epsilon^{1/2}$ splitting in the system's eigenvalues, or are equivalent to perturbing the system's off-diagonal elements. Thus, we need only consider perturbations to the system's off-diagonal coupling matrix elements, as we do in the main text and in the remainder of the appendices. 

First, we consider common-mode resonant frequency perturbations. Such a sensor is described by
\begin{equation}
\begin{split}
    \partial_{t} a &= -i \Omega_0 (1 + \epsilon) a - \gamma a + \gamma b \\
    \partial_{t} b &= -i \Omega_0 (1 + \epsilon) b + \gamma b - \gamma  a.
\end{split}
\end{equation}
Since this perturbation is equivalent to adding a scaled identity matrix to the system's coupling matrix, it only shifts the system's eigenvalue and does not lift the degeneracy. It then has no $\epsilon^{1/2}$ eigenvalue splitting.

For differential resonant frequency perturbations, the sensor is described by
\begin{equation}
\begin{split}
    \partial_{t} a &= -i \Omega_0 (1 + \epsilon) a - \gamma a + \gamma b \\
    \partial_{t} b &= -i \Omega_0 (1 - \epsilon) b + \gamma b - \gamma  a.
\end{split}
\end{equation}
and its normal mode frequencies are given by
\begin{equation}
    \Omega_0 \pm i \sqrt{2 i \gamma \Omega_0 \epsilon - \epsilon^2 \Omega_0^2} \approx \Omega_0 \pm i \sqrt{2 i \gamma \Omega_0 \epsilon}
\end{equation}
where the approximation is that $\epsilon \ll 1$, which is the regime relevant for EP sensing. We see that these eigenfrequencies come in complex-conjugate pairs, with the perturbation encoded in their imaginary parts. If one or both of the imaginary parts of the sensor's eigenfrequencies are positive, the sensor is unstable and saturation effects will return the sensor to a regime where neither eigenfrequency has a positive imaginary component. Once this happens, the sensor's eigenvalues will either no longer be split, or both eigenvalues will have negative imaginary parts, corresponding to the case where both normal modes are lossy and the analysis of \cref{subsec:passiveEpSys,subsec:activeEps} applies.

Having found that perturbing the bare resonant frequencies is not viable for sensing, we consider perturbing the loss and amplification rates. For differential perturbations to these rates, the sensor is described by
\begin{equation}
\begin{split}
    \partial_{t} a &= -i \Omega_0 a - \gamma (1 - \epsilon) a + \gamma b \\
    \partial_{t} b &= -i \Omega_0 b + \gamma (1 + \epsilon) b - \gamma  a.
\end{split}
\end{equation}
and has one eigenfrequency given by
\begin{equation}
    \Omega_0 + i \gamma \epsilon,
\end{equation}
so the perturbation does not lift the degeneracy and shifts the resonance frequency linearly.

Finally, a sensor with a common-mode perturbation to its loss and amplification rates is described by
\begin{equation}\label{app:perturbGainCommon}
\begin{split}
    \partial_{t} a &= -i \Omega_0 a - \gamma (1 - \epsilon) a + \gamma b \\
    \partial_{t} b &= -i \Omega_0 b + \gamma (1 - \epsilon) b - \gamma  a,
\end{split}
\end{equation}
but this perturbation is equivalent to perturbing the coupling between the modes, which we can see by defining $\gamma^\prime \equiv \gamma (1 + \epsilon)$ after which \cref{app:perturbGainCommon} is equivalent to
\begin{equation}
\begin{split}
    \partial_{t} a &= -i \Omega_0 a - \gamma^\prime a + \gamma^\prime (1 + \epsilon) b \\
    \partial_{t} b &= -i \Omega_0 b + \gamma^\prime b - \gamma^\prime (1 + \epsilon) a,
\end{split}
\end{equation}
where we have used the assumption that $\epsilon \ll 1$ to write $1/(1 - \epsilon)$ as $1+\epsilon$.

\section{Solutions of the Classical Equations of Motion}\label{sec:classicalEOM}

\Cref{eq:epQuantumOde} gives the full dynamics of a PT-symmetric EP sensor.
In this section, we solve this equation for the sensor's mean-field dynamics, which are necessary to compute its output frequency spectrum.
Taking the stochastic average of \cref{eq:epQuantumOde} recovers the EP system's classical dynamics given by
\begin{equation}\label{app:ptSymmClassical}
\begin{split}
    \partial_{t} a &= -i \Omega_0 a - \gamma a + \gamma (1 + \bar{\epsilon}) b \\
    \partial_{t} b &= -i \Omega_0 b + \gamma b - \gamma (1 + \bar{\epsilon}) a.
\end{split}
\end{equation}
Taking the Fourier transform we have, 
\begin{equation}\label{app:ptSymmClassicalFt}
\begin{split}
    -i \omega a [\omega] &= -i \Omega_0 a [\omega] - \gamma a [\omega] + \gamma (1 + \bar{\epsilon}) b [\omega] \\
    -i \omega b [\omega] &= -i \Omega_0 b [\omega] + \gamma b [\omega] - \gamma (1 + \bar{\epsilon}) a [\omega].
\end{split}
\end{equation}
and we find the frequency-domain solutions
\begin{equation}\label{laser:eomSolsFreqDomain}
\begin{split}
    a[\omega] =& 2 \pi a_+ \delta[\omega - \Omega_+] + 2 \pi a_- \delta[\omega - \Omega_-] \\
    b[\omega] =& 2 \pi \left(\frac{1-i\sqrt{\bar{\epsilon}(2+\bar{\epsilon})}}{1+\bar{\epsilon}}\right) a_+ \delta[\omega - \Omega_+] + 2 \pi \left(\frac{1+i\sqrt{\bar{\epsilon}(2+\bar{\epsilon})}}{1+\bar{\epsilon}}\right)a_- \delta[\omega - \Omega_-],
\end{split}
\end{equation}
which correspond to the time-domain solutions
\begin{equation}\label{laser:eomSols}
\begin{split}
    a(t) =& \,a_+ e^{-i \Omega_+ t} + a_- e^{-i \Omega_- t} \\
    b(t) =& \left(\frac{1-i\sqrt{\bar{\epsilon}(2+\bar{\epsilon})}}{1+\bar{\epsilon}}\right) a_+ e^{-i \Omega_+ t} + \left(\frac{1+i\sqrt{\bar{\epsilon}(2+\bar{\epsilon})}}{1+\bar{\epsilon}}\right) a_- e^{-i \Omega_- t} \\
    \equiv& \,b_+ e^{-i \Omega_+ t} + b_- e^{-i \Omega_- t}.
\end{split}
\end{equation}
In \cref{laser:eomSolsFreqDomain,laser:eomSols}, $a_\pm$ are complex-constants, and we define the complex-constants $b_\pm$ in the third line of \cref{laser:eomSols} for notational simplicity. 
Due to the spectral poles at $\Omega_\pm$ (see \cref{eq:aOut,eq:qout,eq:pout}), the system will build up macroscopic oscillations at these frequencies. 
When the system reaches steady-state, the values of the constants $a_\pm$ will be determined by nonlinear saturation effects in the sensor's amplifier and loss mechanism. As the sensor's gain and loss mechanisms are properties of the sensor's gain and loss mechanisms and are independent of $\bar{\epsilon}$, $a_\pm$ will be independent of $\bar{\epsilon}$ as well.

To determine the sensor's output frequency spectrum, we need both its phase-quadrature spectrum and its macroscopic output spectrum. We can find the later from \cref{laser:eomSolsFreqDomain} using the boundary condition $a_\text{out}[\omega] = \sqrt{2\gamma} a[\omega]$, we find
\begin{equation}
    a_\text{out} [\omega] = 2\pi \sqrt{2 \gamma} a_+ \delta[\omega - \Omega_+] + 2 \pi \sqrt{2 \gamma} a_- \delta[\omega - \Omega_-]
\end{equation}
Finally, the photon flux $\hat{N}_\text{out}$ in the frequency range $\bar{\omega} - \Delta \omega$ to $\bar{\omega} + \Delta \omega$ of the output field mode $\hat{a}_\text{out}$ is given by \cite{Blow90,Danilishin2012}
\begin{equation}
    \hat{N}_\text{out} (t) = \int_{\bar{\omega} - \Delta \omega}^{\bar{\omega} + \Delta \omega} \frac{d \omega}{2\pi} \int_{\bar{\omega} - \Delta \omega}^{\bar{\omega} + \Delta \omega} \frac{d \omega^\prime}{2\pi} \hat{a}_\text{out}^\dagger[-\omega] \hat{a}_\text{out}[\omega^\prime] e^{i (\omega - \omega^\prime) t},
\end{equation}
so the macroscopic photon flux of out-coupled photons at frequencies $\Omega_\pm$, are given by
\begin{equation}
    N_\pm = 2 \gamma|a_\pm|^2.
\end{equation}

For an oscillator with resonant frequency $\Omega_\text{res}$, the oscillator's frequency spectrum is related to its output phase spectrum by
\begin{equation}
    \bar{S}_{\dot{\varphi}\dot{\varphi}} [\Omega_\text{res} + \Delta \omega] = \Delta \omega^2 \bar{S}_{\varphi \varphi}[\Omega_\text{res} + \Delta\omega]
\end{equation}
and the phase spectrum is in turn related to the output phase quadrature spectrum and the output photon flux, $N$, by \cite{Clerk10,HauTow62}
\begin{equation}
    \bar{S}_{\varphi \varphi}[\Omega_\text{res} + \Delta\omega] = \frac{\Delta\omega^2}{2 N}\bar{S}_{pp}^\text{out}[\Omega_\text{res} + \Delta\omega].
\end{equation}

\section{General Stochastic Solution}\label{sec:noWeakStationary}

\subsection{General Frequency Domain Solution}

In this section, we detail the derivation of \cref{eq:aOut,eq:SppApprox}) from \cref{eq:epCoupledMode}, and show that the conclusions drawn from this equation are unchanged without the simplifying assumptions that we're only interested in the sensor's behavior near resonance and near the EP where $\omega - \Omega_0 \ll \gamma$ and $\epsilon \ll 1$.

Fourier transforming \cref{eq:epCoupledMode}, we find that the fluctuations of the EP sensor are described by
\begin{equation}\label{eq:eomFt}
\begin{split}
    -i \omega \delta \hat{a} [\omega] = -i \Omega_0 \delta \hat{a} [\omega] - \gamma \delta \hat{a} [\omega] + \gamma (1 + \bar{\epsilon}) \delta \hat{b} [\omega] + \gamma b_+ \delta \epsilon [\omega - \Omega_+] + \gamma b_- \delta \epsilon [\omega - \Omega_-] + \sqrt{2\gamma}\delta\hat{a}_\text{in}[\omega] \\
    -i \omega \delta \hat{b} [\omega] = -i \Omega_0 \delta \hat{b} [\omega] + \gamma \delta \hat{b} [\omega] - \gamma (1 + \bar{\epsilon}) \delta \hat{a} [\omega] - \gamma a_+ \delta \epsilon [\omega - \Omega_+] - \gamma a_- \delta \epsilon [\omega - \Omega_-] + \sqrt{2\gamma}\delta\hat{b}_\text{amp}^\dagger[\omega]
\end{split}
\end{equation}
where we have used the mean time evolution of modes $a$ and $b$ given by \cref{laser:eomSols}.
Using the boundary condition $\delta \hat{a}_\text{out} = \sqrt{2 \gamma} \, \delta \hat{a} - \delta\hat{a}_\text{in}$ we find that the fluctuations in the EP sensor's output mode are related to the fluctuations in $\epsilon$ and in its noise modes by
\begin{equation}\label{eq:aOutFt}
\begin{split}
    \delta \hat{a}_\text{out} [\omega] =& \frac{-2 \gamma^2 (1+\epsilon) \delta \hat{b}_\text{amp}^\dagger[\omega] + 2 \left(\gamma^2 + i \gamma (\omega - \Omega_0) \right) \delta \hat{a}_\text{in}[\omega]}{(\omega - \Omega_-)(\omega - \Omega_+)} - \delta \hat{a}_\text{in}[\omega] \\
    &+ \frac{\sqrt{2}\gamma^{3/2} \Big( (1+\epsilon)a_- \gamma + b_- (\gamma + i(\omega - \Omega_0)) \Big) }{(\omega - \Omega_-)(\omega - \Omega_+)}\delta\epsilon [\omega - \Omega_-] \\
    &- \frac{\sqrt{2}\gamma^{3/2} \Big( (1+\epsilon)a_+ \gamma + b_+ (\gamma + i(\omega - \Omega_0)) \Big) }{(\omega - \Omega_-)(\omega - \Omega_+)} \delta\epsilon [\omega - \Omega_+].
\end{split}
\end{equation}
There is some subtlety in converting this equation into a relation between quadrature operators. Since the system's bare resonance frequencies, $\Omega_0$ are set by time delays in the physical coupled-mode system, terms containing $\Omega_0$ originate from $e^{i \Omega_0 \tau}$ terms in the system's full non-Markovian dynamics, where $\tau$ is the feedback delay time [see \cref{sec:nonMarkov}]. The resonance frequencies are set such that $e^{i \Omega_0 \tau} = 1$, so $\Omega_0$ terms drop out of the full dynamics. In \cref{eq:eomFt}, these terms appear to enforce a particular oscillation mode in the sensor, but we must recall that these terms originate from time delays in the physical system under consideration. Thus, when we take the adjoint of \cref{eq:aOutFt} we should either neglect the $\Omega_0$ terms and then take $\omega \rightarrow \omega - \Omega_0$ in our final expression, or we should flip the sign of $\Omega_0$ whenever we flip the sign of $\omega$, i.e. if $f[\omega] = \omega - \Omega_0$, $f[-\omega] = - (\omega - \Omega_0)$.

With this caveat about the treatment of $\Omega_0$ in mind, we find that the output quadratures are related to $\delta \epsilon$ and the quadratures of the noise modes by
\begin{equation}\label{eq:qout}
\begin{split}
    \delta \hat{q}_\text{out}[\omega] =& \left( \frac{2\gamma(\gamma + i (\omega - \Omega_0))}{(\omega - \Omega_-)(\omega - \Omega_+)} - 1 \right) \delta \hat{q}_\text{in}[\omega] - \frac{2 (1 + \epsilon)\gamma^2}{(\omega - \Omega_-)(\omega - \Omega_+)} \delta \hat{q}_\text{amp}[\omega] \\
    & + \frac{\sqrt{2}\gamma^{3/2} \Big( (1 + \epsilon)q_{a}^-\gamma + q_b^- (\gamma + i (\omega - \Omega_0)) \Big)}{(\omega - \Omega_-)(\omega - \Omega_+)} \delta \epsilon [\omega - \Omega_-] \\
    & + \frac{\sqrt{2}\gamma^{3/2} \Big( (1 + \epsilon)q_a^+\gamma + q_b^+ (\gamma + i (\omega - \Omega_0)) \Big)}{(\omega - \Omega_-)(\omega - \Omega_+)} \delta \epsilon [\omega - \Omega_+]
\end{split}
\end{equation}
and
\begin{equation}\label{eq:pout}
\begin{split}
    \delta \hat{p}_\text{out}[\omega] =& \left( \frac{2\gamma(\gamma + i (\omega - \Omega_0))}{(\omega - \Omega_-)(\omega - \Omega_+)} - 1 \right) \delta \hat{p}_\text{in}[\omega] + \frac{2 (1 + \epsilon)\gamma^2}{(\omega - \Omega_-)(\omega - \Omega_+)} \delta \hat{p}_\text{amp}[\omega] \\
    & + \frac{\sqrt{2}\gamma^{3/2} \Big( (1 + \epsilon)p_a^-\gamma + p_b^- (\gamma + i (\omega - \Omega_0)) \Big)}{(\omega - \Omega_-)(\omega - \Omega_+)} \delta \epsilon [\omega - \Omega_-] \\
    & + \frac{\sqrt{2}\gamma^{3/2} \Big( (1 + \epsilon)p_a^+\gamma + p_b^+ (\gamma + i (\omega - \Omega_0)) \Big)}{(\omega - \Omega_-)(\omega - \Omega_+)} \delta \epsilon [\omega - \Omega_+],
\end{split}
\end{equation}
where we have defined $q_{a,b}^\pm$ and $p_{a,b}^\pm$ by to be the amplitude and phase quadratures of $a_\pm$ and $b_\pm$.

\subsection{Weak Force Sensing}

In the main text, we saw that EP sensors have no advantage in weak force sensing in the limit that these sensors operate near the EP and near resonance and that $\delta \epsilon$ is weak-stationary. From \cref{eq:qout,eq:pout}, we see that these sensors also have no fundamental sensing advantage without these assumptions. The output quadratures' response to deviations in $\epsilon$ is enhanced when the denominator $(\omega - \Omega_-)(\omega - \Omega_+)$ is small, but when this is the case, the noise terms from $\delta q_\text{in}$, $\delta p_\text{in}$, $\delta q_\text{amp}$, and $\delta p_\text{amp}$ are enhanced by an equal amount, so there is no overall sensing enhancement, even in the absence of the simplifying assumptions made in the main text.

\subsection{Parameter Estimation}

Instead of measuring a weak force coupled to $\delta \epsilon$, we can consider using the EP sensor to estimate the mean value of $\epsilon$, $\bar{\epsilon}$. In this case, we need not consider fluctuations in $\epsilon$ and can neglect the final two terms in \cref{eq:qout,eq:pout}. 

Changes in $\bar{\epsilon}$ will shift the EP sensor's resonant frequencies, and our ability to measure these frequency shifts will be limited by the sensor's output frequency noise, which is related to its output phase quadrature spectrum. From \cref{eq:pout}, and assuming that the various noise modes are uncorrelated with each other, we find that this spectrum is given by
\begin{equation}\label{eq:Sppout}
\begin{split}
    \bar{S}_{pp}^\text{out}[\omega] =& \left( \frac{4 \gamma^4}{(\omega - \Omega_-)^2(\omega - \Omega_+)^2} - \frac{4 \gamma^2}{(\omega - \Omega_-)(\omega - \Omega_+)} + 1 \right) \bar{S}_{pp}^\text{in}[\omega] + \left( \frac{4 (1 + \epsilon)^2 \gamma^4}{(\omega - \Omega_-)^2(\omega - \Omega_+)^2} \right) \bar{S}_{pp}^\text{amp}[\omega].
\end{split}
\end{equation}
To determine the EP sensor's linewidth and frequency fluctuations, we want to consider its phase quadrature spectrum near its resonant frequencies $\Omega_\pm$, to do so, we define an offset frequency from resonance, $\Delta\omega_\pm$ by $\Delta\omega_\pm \equiv \omega - \Omega_\pm$. We also consider the sensor near the EP such that $\bar{\epsilon} \ll 1$. Assuming that $\Delta \omega_\pm \ll 2\sqrt{2 \bar{\epsilon}}$ and working to leading order in $\Delta\Omega_\pm$ and $\bar{\epsilon}$, we find
\begin{equation}
    \bar{S}_{pp}^\text{out}[\omega] = \frac{\gamma^2}{2 \bar{\epsilon} \Delta \omega_\pm^2} \left(\bar{S}_{pp}^\text{in}[\omega] + \bar{S}_{pp}^\text{amp}[\omega]\right)
\end{equation}

Assuming that the noise modes are in thermal states with mean photon occupation numbers $n_\text{amp}$ and $n_\text{in}$, the output phase quadrature fluctuations are
\begin{equation}\label{eq:outputPhaseQuadSpec}
    \bar{S}_{pp}^\text{out}[\omega] 
    = \frac{\gamma^2}{2 \bar{\epsilon} \Delta \omega_\pm^2} (1 + n_\text{amp} + n_\text{in}) 
    = \frac{\gamma^2}{2 \bar{\epsilon} \Delta \omega_\pm^2} \left(1 + 2\bar{n}_\text{th} \right) 
\end{equation}
where $\bar{n}_\text{th} \equiv (n_\text{amp} + n_\text{in})/2$ is the mean thermal occupation number of the in-coupled mode and the amplifier. This expression is the same as \cref{eq:SppApprox}, but in the derivation of \cref{eq:outputPhaseQuadSpec} we did not make any approximations until after deriving an exact form for the sensor's output phase quadrature spectrum.

\section{Non-Markovian PT-Symmetric Sensor Dynamics}\label{sec:nonMarkov}

In the main text, we primarily considered PT-symmetric sensors in the Markovian limit where the resonator modes $\hat{a}$ and $\hat{b}$ are memoryless. In a physical EP sensor, depicted in \cref{fig:epSensorSchematic} and in more detail in \cref{fig:fbOscEp} below, the resonator modes will have memory and non-Markovian dynamics. For instance, in the case of an optical PT-symmetric sensor, the propagation time of light around an optical cavity causes the system to have a memory and experience non-Markovian dynamics. In this section, we show that a non-Markovian PT-symmetric sensor has a signal to noise ratio (SNR) with the same scaling with perturbation strength, $\epsilon$ as a non-Markovian sensor. The only difference between the two is a small prefactor correction accounting for non-Markovanity.

We will first examine the coupled-mode equations for a non-Markovian PT-symmetric EP sensor and show that they reduce to the Markovian dynamics discussed in the main text in the appropriate limit. We will then examine the signal and noise properties of these non-Markovian sensors and show that they only correct the Markovian results by an order-unity prefactor that accounts for inexact Markovanity.

\begin{figure}[t!]
    \centering
    \includegraphics[width=0.4\columnwidth]{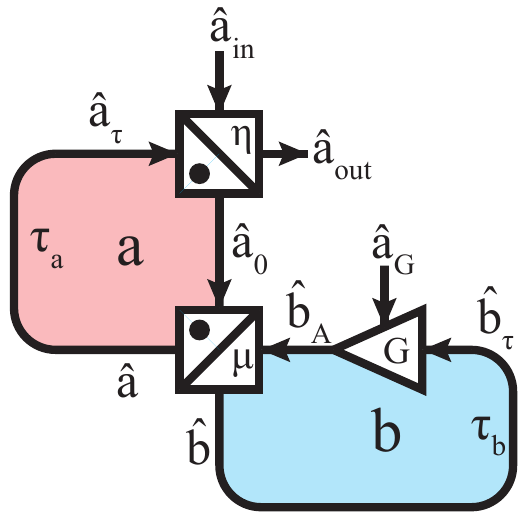}
    \caption{\label{fig:fbOscEp} An abstract model of a realistic EP sensor. This model explicitly accounts for feedback delays present in any realistic implementation of an EP sensor.}
\end{figure}

\subsection{The Relation between Markovian and Non-Markovian EP Sensors}\label{sec:deriveCoupledMode}

We will now show how the equations governing the coupled modes in \cref{fig:fbOscEp} reduce to the Markovian equations of motion given by \cref{eq:epOde}. 
In this non-Markovian model, we have two cavity modes $\hat{a}$ and $\hat{b}$, that are coupled via a beamsplitter interaction and fed back into their respective cavities with time delays $\tau_a$ and $\tau_b$. 
Mode $\hat{a}$ is lossy, and is partially out-coupled via a beam splitter with amplitude transmissivity $\sqrt{\eta}$. 
Mode $\hat{b}$ is amplified and experiences an amplitude gain of $\sqrt{1+G}$ after a single pass through its gain medium. 
These modes are coupled by a second beamsplitter, with amplitude transmissivity $\sqrt{\mu}$. 
For a PT-symmetric sensor, the modes $\hat{a}$ and $\hat{b}$ must be symmetric, so we require $\tau_a = \tau_b \equiv \tau$. Such a system is depicted in \cref{fig:fbOscEp}. 

In the limit that $\eta \rightarrow 0$ and $\tau \rightarrow 0$ with $\eta/(2\tau)$ held constant, the dynamics of this non-Markovian sensor reduce to those of \cref{eq:epOde}, as will be shown in detail below.

With these PT-symmetric requirements in mind, the equations of motion describing this system are given in the time domain and Heisenburg picture by
\begin{equation}\label{eq:loopEqEp}
\begin{split}
    \hat{a}_0 (t) &= \sqrt{\eta} \, \hat{a}_\text{in} (t) - \sqrt{1 - \eta} \, \hat{a}_\tau (t) \\
    \hat{a}_\text{out} (t) &= \sqrt{\eta} \, \hat{a}_\tau (t) + \sqrt{1 - \eta} \, \hat{a}_\text{in} (t) \\
    \hat{b} (t) &= \sqrt{\mu} \, \hat{a}_0 (t) + \sqrt{1 - \mu} \, \hat{b}_\text{A} (t) \\
    \hat{a} (t) &= \sqrt{\mu} \, \hat{b}_\text{A} (t) - \sqrt{1 - \mu} \, \hat{a}_0 (t) \\
    \hat{a}_\tau (t) &= \hat{a} (t - \tau) \\
    \hat{b}_\tau (t) &= \hat{b} (t - \tau) \\
    \hat{b}_\text{A} (t) &= \sqrt{1 + G} \, \hat{b}_\tau (t) + \sqrt{G}\, \hat{b}_\text{amp}^\dagger (t)
\end{split}
\end{equation}

We will first show how \cref{eq:loopEqEp} reduce to \cref{eq:epOde} in the appropriate limit. Since \cref{eq:epOde} is a set of equations for a classical system, we will focus on the mean dynamics and take $a_\text{in} = b_\text{amp} = 0$, where we decompose an operator $\hat{A}$ into its mean value and quantum fluctuations via $\hat{A} = A + \delta \hat{A}$ where $\langle \delta \hat{A} \rangle = 0$. Making this substitution and eliminating $a_\tau$, $a_0$, $b_\tau$, and $b_\t{A}$ from the set of equations, we have
\begin{equation}\label{eq:abDiscreteTimeEvol}
\begin{split}
    a(t + \tau) &= \sqrt{1 - \mu} \sqrt{1 - \eta} \,a(t) + \sqrt{\mu} \sqrt{1 + G} \, b(t) \\
    b(t + \tau) &= -\sqrt{\mu}\sqrt{1 - \eta} \, a(t) + \sqrt{1 - \mu} \sqrt{1 + G} \, b(t)
\end{split}
\end{equation}
From here, we take $a$ and $b$ to have the form of Fourier series with time-dependent coefficients, namely
\begin{equation}\label{eq:abFourierDecomp}
\begin{split}
    a(t) = \sum_{n = - \infty}^\infty c_n (t) e^{2 \pi i n t / \tau} \\
    b(t) = \sum_{n = - \infty}^\infty d_n (t) e^{2 \pi i n t / \tau}
\end{split}
\end{equation}
where $c_n$ and $d_n$ vary slowly relative to the exponential terms.
Plugging these decompositions of $a$ and $b$ into \cref{eq:abDiscreteTimeEvol}, we have
\begin{equation}
\begin{split}
    c_n(t + \tau) &= \sqrt{1 - \mu} \sqrt{1 - \eta} \,c_n(t) + \sqrt{\mu} \sqrt{1 + G} \, d_n(t) \\
    d_n(t + \tau) &= -\sqrt{\mu}\sqrt{1 - \eta} \, c_n(t) + \sqrt{1 - \mu} \sqrt{1 + G} \, d_n(t)
\end{split}
\end{equation}
which are nearly the same as \cref{eq:abDiscreteTimeEvol}, but now the periodicity has been lopped into the exponential terms in \cref{eq:abFourierDecomp}. Assuming that $\eta$, $\mu$ and $G$ are all small such that $c_n$ and $d_n$ do not change much over one period, we can approximate $c_n(t + \tau)$ ($d_n(t + \tau)$) by $c_n(t) + \tau \partial_t c_n (t)$ ($d_n(t) + \tau \partial_t d_n (t)$). We then have
\begin{equation}\label{eq:diffEqFbOsc1}
\begin{split}
    \partial_t c_n (t) &\approx \left(\frac{\sqrt{1 - \mu} \sqrt{1 - \eta} - 1}{\tau} \right) c_n(t) + \left( \frac{\sqrt{\mu} \sqrt{1 + G}}{\tau} \right) d_n(t) \\
    \partial_t d_n (t) &\approx -\left(\frac{\sqrt{\mu}\sqrt{1 - \eta}}{\tau} \right) c_n(t) + \left(\frac{\sqrt{1 - \mu} \sqrt{1 + G}-1}{\tau} \right) d_n(t)
\end{split}
\end{equation}
We now take $G = \eta$ and $\mu = (1/4)(1 + \epsilon)^2 \eta^2$. To first order in $\eta$, \cref{eq:diffEqFbOsc1} reduces to
\begin{equation}\label{eq:diffEqFbOsc2}
\begin{split}
    \partial_t c_n (t) &\approx -\left(\frac{\eta}{2 \tau} \right) c_n(t) + \left( \frac{\eta (1 + \epsilon)}{2 \tau} \right) d_n(t) \\
    \partial_t d_n (t) &\approx -\left(\frac{\eta (1 + \epsilon)}{2 \tau} \right) c_n(t) + \left(\frac{\eta}{2 \tau} \right) d_n(t)
\end{split}
\end{equation}
Taking $\eta \rightarrow 0$ with $\eta/(2 \tau) \equiv \gamma$ held constant, the higher order terms vanish, and we have
\begin{equation}\label{eq:diffEqFbOsc3}
\begin{split}
    \partial_t c_n (t) &= - \gamma \, c_n(t) + \gamma (1 + \epsilon) \, d_n(t) \\
    \partial_t d_n (t) &= - \gamma (1 + \epsilon) \, c_n(t) + \gamma \, d_n(t)
\end{split}
\end{equation}
Assuming that modes $a$ and $b$ only resonate near one frequency, $\Omega_0$, we have
\begin{equation}\label{eq:diffEqFbOsc4}
\begin{split}
    \partial_t a (t) &= - i \Omega_0 a(t) - \gamma a(t) + \gamma (1 + \epsilon) b(t) \\
    \partial_t b (t) &= - i \Omega_0 b(t) + \gamma b(t) - \gamma (1 + \epsilon) a(t)
\end{split}
\end{equation}
which is the same as \cref{eq:epOde}.

In conclusion, we have seen that the dynamics of \cref{eq:loopEqEp} reduce to those of \cref{eq:epOde} if we take
\begin{equation}\label{eq:gMuParamVals}
\begin{split}
    G &= \eta + \mathcal{O}(\eta^2) \\
    \mu &= \frac{(1 + \epsilon)^2 \eta^2}{4} + \mathcal{O}(\eta^3)
\end{split}
\end{equation}
and then take the limit of $\eta \rightarrow 0$ with $\eta / (2 \tau) = \gamma$ held constant. In practice, the coupling beam splitter's transmissivity, $\mu$ can be linear in $\epsilon$ with $\mu = \frac{(1 + 2\epsilon) \eta^2}{4}$ since $\epsilon \ll 1$ for EP sensors, so we can linearize \cref{eq:gMuParamVals} about $\epsilon = 0$.

\subsection{Signal and Noise Properties of a Non-Markovian EP Sensor}

We have seen in the previous section that the non-Markovian system described by \cref{eq:loopEqEp} describes the same dynamics as the Markovian coupled mode system of \cref{eq:epQuantumOde} if \cref{eq:gMuParamVals} is satisfied and we take the limit $\eta, \tau \rightarrow 0$ with $\eta/(2 \tau) = \gamma$. We now examine the noise properties of this non-Markovian system.

\Cref{eq:gMuParamVals} only gives $G$ and $\mu$ to leading order in $\eta$, but we want these values to all orders in $\eta$ to determine the non-Markovian corrections to the sensor's signal and noise properties. The value of $G$ is fixed by gain-loss balance, which requires that ``round-trip-gain''$\times$``rount-trip-loss''$=1$, i.e. $\sqrt{1+G}\sqrt{1-\eta}=1$. The value of $\mu$ is set by the requirement that the sensor have a resonance at the EP frequencies.\footnote{To recover the dynamics of the Markovian EP sensor in the appropriate limit, the non-Markovian system's transfer functions from the input and amplifier mode to the output mode should have first order poles at $\Omega_0 \pm (\eta/(2 \tau))\sqrt{\epsilon (1 + \epsilon)}$. This requirement determines the expression for $\mu$ in \cref{eq:GandMuEP}.}
Together, these require
\begin{equation}\label{eq:GandMuEP}
\begin{split}
    G &= \frac{\eta}{1-\eta} \\
    \mu &= 1 - 4 \left[\frac{\cos \left(\frac{\eta}{2}\sqrt{\epsilon(2 + \epsilon)}\right)}{\sqrt{1 - \eta}+ 1/\sqrt{1-\eta}} \right]^2
\end{split}
\end{equation}
which are equivalent to \cref{eq:gMuParamVals} to leading order in $\eta$ and specify the higher-order correction terms that vanish in the $\eta,\tau \rightarrow 0$ limit.

To evaluate the signal and noise properties of the sensor, we analyze the Fourier transform of \cref{eq:loopEqEp},
\begin{equation}\label{eq:loopEqEpFt}
\begin{split}
    \hat{a}_0 [\omega] &= \sqrt{\eta} \, \hat{a}_\text{in} [\omega] - \sqrt{1 - \eta} \, \hat{a}_\tau [\omega] \\
    \hat{a}_\text{out} [\omega] &= \sqrt{\eta} \, \hat{a}_\tau [\omega] + \sqrt{1 - \eta} \, \hat{a}_\text{in} [\omega] \\
    \hat{b} [\omega] &= \sqrt{\mu} \, \hat{a}_0 [\omega] + \sqrt{1 - \mu} \, \hat{b}_\text{A} [\omega] \\
    \hat{a} [\omega] &= \sqrt{\mu} \, \hat{b}_\text{A} [\omega] - \sqrt{1 - \mu} \, \hat{a}_0 [\omega] \\
    \hat{a}_\tau [\omega] &= e^{i \omega \tau} \hat{a} [\omega] \\
    \hat{b}_\tau [\omega] &= e^{i \omega \tau} \hat{b} [\omega] \\
    \hat{b}_A [\omega] &= \sqrt{1 + G} \, \hat{b}_\tau [\omega] + \sqrt{G}\, \hat{b}_\text{amp}^\dagger [\omega].
\end{split}
\end{equation}
We solve these equations for $\hat{a}_\text{out} [\omega]$ in terms of $\hat{a}_\text{in} [\omega]$ and $\hat{b}_\text{amp}^\dagger [\omega]$ and find
\begin{equation}\label{eq:markovTfs}
    \hat{a}_\text{out} [\omega] = H_\text{in} [\omega] \hat{a}_\text{in} [\omega] + H_\text{amp}[\omega] \hat{b}_\text{amp}^\dagger [\omega],
\end{equation}
where the transfer functions $H_\text{in}$ and $H_\text{amp}$ are given by
\begin{equation}\label{eq:tfs}
\begin{split}
    H_\text{in} [\omega] &= \frac{e^{-i \omega \tau} \left[2 + (2 - \eta) e^{2 i \omega \tau} - 3\eta + \eta^2 - 4 e^{i \omega\tau} (1-\eta) \cos \left( \frac{\eta}{2}\sqrt{\epsilon(2 + \epsilon)} \right) \right]}{2 \sqrt{1 - \eta} (2 - \eta) \left[\cos(\omega \tau) - \cos\left( \frac{\eta}{2}\sqrt{\epsilon(2 + \epsilon)} \right) \right]}\\
    H_\text{amp} [\omega] &= \frac{\eta \sqrt{2 - \eta(2 - \eta) - 2 (1-\eta)\cos\left(\eta \sqrt{\epsilon (2 + \epsilon )}\right)}}{2 \sqrt{1 - \eta} (2 - \eta) \left[\cos(\omega \tau) - \cos\left( \frac{\eta}{2}\sqrt{\epsilon(2 + \epsilon)} \right) \right]}.
\end{split}
\end{equation}
These transfer functions have resonances where $\omega = \Omega_0 \pm\eta\sqrt{\epsilon(2 + \epsilon)}/(2\tau)$, which are the eigenfrequencies of the Markovian EP sensor. Here, we have made use of the assumption that $\Omega_0 \tau = 2 \pi n$ for $n \in \mathbb{Z}$, i.e. $\Omega_0$ is a resonance of the uncoupled system.

To compare the non-Markovian sensor with the Markovian sensor, we consider the transfer functions close to resonance. Let $\omega = \Omega_\pm + \Delta \omega_\pm \equiv \Omega_0 \pm \eta\sqrt{\epsilon(2 + \epsilon)}/(2\tau) + \Delta \omega_\pm$ such that $\Delta \omega_\pm$ denotes an offset frequency from resonance. To leading order in $\Delta \omega_\pm$ and $\epsilon$, the transfer functions are given by
\begin{equation}
\begin{split}
    H_\text{in} [\omega] &\approx \mp \frac{\eta}{\sqrt{2 - 2\eta}(2 - \eta)\tau \sqrt{\epsilon} \, \Delta \omega_\pm} \\
    H_\text{amp} [\omega] &\approx \mp \frac{\eta}{\sqrt{2 - 2\eta}(2 - \eta)\tau \sqrt{\epsilon} \, \Delta\omega_\pm}.
\end{split}
\end{equation}
We observe that these transfer functions are equal in this limit and scale as $1/\sqrt{\epsilon}$, which will cause the same frequency noise scaling with $\epsilon$ as in the Markovian case. In the Markovian limit that $\eta,\tau \rightarrow 0$ with $\eta/(2 \tau) = \gamma$, we find 
\begin{equation}
    H_\text{in} [\omega] = H_\text{amp} [\omega] \approx \mp \frac{\gamma}{\sqrt{2 \epsilon} \, \Delta \omega_\pm}.
\end{equation}

Using \cref{eq:markovTfs}, we find that the output phase quadrature spectrum is related to the input and amplifier phase quadrature spectra by
\begin{equation}
    \bar{S}_{pp}^\text{out}[\omega] = |H_\text{in} [\omega]|^2 \bar{S}_{pp}^\text{in}[\omega] + |H_\text{amp} [\omega]|^2 \bar{S}_{pp}^\text{amp}[\omega]
\end{equation}
as long as the input and amplifier modes are uncorrelated.
In the Markovian limit, we find that the system's output phase quadrature spectrum is given by
\begin{equation}\label{eq:SppMarkov}
    \bar{S}_{pp}^\text{out}[\omega] = \frac{\gamma^2 (1 + 2 \bar{n}_\text{th})}{2 \epsilon \Delta \omega_\pm^2},
\end{equation}
for input and amplifier modes in thermal states such that $\bar{S}_{pp}^\text{in}[\omega] = \bar{S}_{pp}^\text{amp}[\omega] = 1 + 2 \bar{n}_\text{th}$.
This equation is in agreement with \cref{eq:SppApprox}, which was derived via the input-output formalism of refs. \cite{Collett84,Gardiner85}.

In the non-Markovian case, we find that the output phase quadrature spectrum is given by
\begin{equation}\label{eq:SppNonMarkov}
    \bar{S}_{pp}^\text{out}[\omega] = \frac{\eta^2 (1 + 2 \bar{n}_\text{th})}{2(1-\eta)(2-\eta)^2 \tau^2 \epsilon \Delta \omega_\pm^2},
\end{equation}
which gives a small correction to the prefactor of \cref{eq:SppMarkov} coming from higher order terms in $\eta$, but has the same scaling with $\epsilon$. While $\eta$ and $\tau$ are not infinitesimally small in physical systems, we must have $\eta \ll 1$ to consider modes $\hat{a}$ and $\hat{b}$ to be resonant modes.

From \cref{eq:SppNonMarkov}, we see that non-Markovian PT-symmetric EP sensors suffer from the same frequency noise scaling as Markovian PT-symmetric EP sensors. Both sensors have frequency noise power spectra that scale as $1/\epsilon$, which exactly cancels the increase in sensitivity from their resonant frequency splitting. Thus, non-Markovian PT-symmetric EP sensors so not achieve any better SNR than Markovian PT-symmetric EP sensors or traditional sensors operated far from EPs.

\section{A PT-Symmetric EP Sensor with a Phase-Sensitive Amplifier}\label{sec:phaseSensitveSensor}

In this section, we consider the SNR of a PT-symmetric sensor based on a phase-sensitive amplifier instead of a more traditional phase-insensitive amplifier as was considered above. We have seen that while operating near a PT-symmetric EP increases a sensor's signal, it also increases its noise such that PT-symmetric sensors offer no improvement in fundamental performance beyond that of traditional sensors. It then seems plausible that by using a noiseless phase-sensitive amplifier, it may be possible to retain the improved sensitivity to small perturbations while eliminating the excess noise. This section fills in the mathematical details of the phase-sensitive amplifier discussed in the main text.

As discussed in the main text, we find that even in the presence of loss, PT-symmetric sensors with purely phase-sensitive amplifiers, amplifiers that are equivalent to ideal single-mode squeezers, can achieve fundamentally enhanced SNR. While this result is robust against loss, it is not robust against phase-noise, which will likely limit these sensor's use in practice. We find that non-Markovian dynamics provide an order-unity prefactor correction to the Markovian results, mirroring the results for sensors with phase-insensitive amplifiers discussed in \cref{sec:nonMarkov}.

\subsection{Markovian Dynamics}\label{sec:phSensMarkovDynamics}

We first consider a PT-symmetric EP sensor where the amplified mode interacts with a phase-sensitive amplifier in the Markovian limit. First, we derive \cref{eq:phSensLangevain} from a Hamiltonian model. We consider a three-mode system consisting of modes $\hat{a}$, $\hat{b}$, and a pump mode $\hat{c}$ with twice the frequency of modes $\hat{a}$ and $\hat{b}$. We assume that modes $\hat{a}$ and $\hat{b}$ are coupled through a beam-splitter-like interaction and that modes $\hat{b}$ and $\hat{c}$ interact through a second-order nonlinear process. This system can be described by the Hamiltonian
\begin{equation}\label{eq:phSensHamiltonian}
    \hat{H} = \hbar \Omega_0 \hat{a}^\dagger \hat{a} + \hbar \Omega_0 \hat{b}^\dagger \hat{b} + 2\hbar \Omega_0 \hat{c}^\dagger \hat{c} + i \hbar (1 + \epsilon) \gamma \left( \hat{a}^\dagger \hat{b} - \hat{b}^\dagger \hat{a} \right) + \frac{i \hbar \kappa}{2} \left( \hat{b}^{\dagger \, 2} \hat{c} - \hat{b}^2 \hat{c}^\dagger \right)
\end{equation}
with $\{\gamma, \kappa\} \in \mathbb{R}$.
The closed-system dynamics of an operator $\hat{X}$ evolving under a Hamiltonian $\hat{H}$ are given by
\begin{equation}
    \partial_t\hat{X} = - \frac{i}{\hbar} [\hat{X}, \hat{H}],
\end{equation}
so the system's mode evolution under the Hamiltonian given in \cref{eq:phSensHamiltonian} is
\begin{equation}
\begin{split}
    \partial_t \hat{a} &= -i \Omega_0 \hat{a} + (1+\epsilon) \gamma \hat{b} \\
    \partial_t \hat{b} &= -i \Omega_0 \hat{b} + \kappa \hat{b}^\dagger \hat{c} - (1+\epsilon) \gamma \hat{a} \\
    \partial_t \hat{c} &= -2 i \Omega_0 \hat{c} - \frac{\kappa}{2} \hat{b}^2
\end{split}
\end{equation}
Assuming the pump mode, $\hat{c}$ is in a macroscopic coherent state which is undepleted by the parametric interaction with the amplified mode, $\hat{b}$, we can neglect the $- \frac{\kappa}{2} \hat{b}^2$ term in the pump mode's evolution. With these assumptions, the evolution of the pump mode is approximately given by
\begin{equation}
    \hat{c}(t) \approx c_0 e^{-2i\Omega_0 t},
\end{equation}
where $c_0 \in \mathbb{C}$ is a complex constant determined by the pump field's amplitude and phase. Using this expression for the pump field, the evolution of the field modes $\hat{a}$ and $\hat{b}$ is given by
\begin{equation}
\begin{split}
    \partial_t \hat{a} &= -i \Omega_0 \hat{a} + (1+\epsilon) \gamma \hat{b} \\
    \partial_t \hat{b} &= -i \Omega_0 \hat{b} + e^{-2 i \Omega_0 t} r \hat{b}^\dagger - (1+\epsilon) \gamma \hat{a},
\end{split}
\end{equation}
where we have defined $r \equiv c_0 \kappa$, which is the rate of phase-sensitive amplification.

Having derived the closed-system dynamics, we now add gain and loss terms to fully describe the system.
We assume that the amplified mode, $\hat{b}$, has a loss rate $\gamma_b$, which allows us to consider how robust these results are against loss. While we label the out-coupling rate of modes $\hat{a}$ and $\hat{b}$ by $\gamma_a$ and $\gamma_b$, we assume that we can observe the out-coupled field from mode $\hat{a}$ while the out-coupled field from mode $\hat{b}$ is lost to the environment. The phase-sensitive PT-symmetric EP sensor has its full quantum dynamics governed by \cite{Collett84}
\begin{equation}\label{eq:phSensCoupledModeFull}
\begin{split}
    \partial_{t} \hat{a} &= -i \Omega_0 \hat{a} - \gamma_a \hat{a} + \gamma_a (1 + \bar{\epsilon} + \delta \epsilon) \hat{b} + \sqrt{2 \gamma_a}\, \delta\hat{a}_\text{in} \\
    \partial_{t} \hat{b} &= -i \Omega_0 \hat{b} + \gamma_\text{amp} \hat{b} + e^{-2i\Omega_0 t} r \hat{b}^\dagger - \gamma_b \hat{b} - \gamma_a (1 + \bar{\epsilon} + \delta \epsilon) \hat{a} + \sqrt{2 \gamma_\text{amp}}\, \delta\hat{b}_\text{amp}^\dagger + \sqrt{2 \gamma_a} \, \delta\hat{b}_\text{in}
\end{split}
\end{equation}
where $\gamma_\t{amp}$ is the rate of phase-insensitive amplification for mode $\hat{b}$ and the mode $\delta\hat{b}_\text{in}$ is a noise mode introduced by a loss rate $\gamma_b$ in the sensor's amplified mode. 
Since we want the oscillator to resonate at the same frequencies as the oscillator with no phase-sensitive amplification, the uncoupled amplification rate of mode $\hat{b}$'s amplitude quadrature, $\hat{q}_b$ must be $\gamma_a$, which fixes $\gamma_\t{amp} = \gamma_a + \gamma_b - r$.

\subsubsection{Mean Dynamics}

As in the phase-insensitive case, we take a statistical average over the sensor's equations of motion to find its mean evolution. To do so, we take the stochastic average of \cref{eq:phSensCoupledModeFull} and then go to a rotating frame by defining
\begin{equation}
\begin{split}
    \hat{A} &= e^{i\Omega_0 t}\hat{a} \\
    \hat{B} &= e^{i\Omega_0 t}\hat{b}.
\end{split}
\end{equation}
\Cref{eq:phSensCoupledModeFull} then reduces to
\begin{equation}\label{eq:phSensMeanRotFrame}
\begin{split}
    \partial_{t} A &= - \gamma_a A + \gamma_a (1 + \bar{\epsilon}) B \\
    \partial_{t} B &= (\gamma_a - r) B + r B^\dagger - \gamma_a (1 + \bar{\epsilon}) A
\end{split}
\end{equation}
Unlike the phase-insensitive case, where we could work out the sensor's mean dynamics in terms of the lowering operators $\hat{a}$ and $\hat{b}$, the mean dynamics of the phase-sensitive EP sensor are more simply evaluated in the terms of the quadratures as
\begin{equation}
\begin{split}
    \partial_{t} q_A &= - \gamma_a q_A + \gamma_a (1 + \bar{\epsilon}) q_B \\
    \partial_{t} q_B &= \gamma_a q_B - \gamma_a (1 + \bar{\epsilon}) q_A \\
    \\
    \partial_{t} p_A &= - \gamma_a p_A + \gamma_a (1 + \bar{\epsilon}) p_B \\
    \partial_{t} p_B &= (\gamma_a - 2r)p_B - \gamma_a (1 + \bar{\epsilon}) p_A
\end{split}
\end{equation}
We find that the system's eigenfrequencies are given by
\begin{equation}
    \pm i\gamma_a \sqrt{\bar{\epsilon}(2 + \bar{\epsilon})}, \quad -r \pm i\sqrt{(r + \bar{\epsilon}\gamma_a)(\gamma_a(2 + \bar{\epsilon}) - r)}
\end{equation}
where the first pair of eigenvalues correspond to the evolution of the system's amplitude quadratures, and the second pair to its phase quadratures. Due to the negative real part of the phase quadrature's eigenvalues, the system's phase-quadrature amplitudes decay over time and are zero in steady-state.

The sensor's steady-state, mean-field behavior is then determined by
\begin{equation}\label{eq:phsensMeanDynamics}
\begin{split}
    q_a(t) &= q_0 \left(e^{i \Omega_- t} + e^{-i \Omega_+ t} \right) + q_0^\dagger \left(e^{-i \Omega_- t} + e^{i \Omega_+ t} \right) \\
    q_b(t) &= \left(\frac{1 - i \sqrt{\bar{\epsilon}(2 + \bar{\epsilon})}}{1 + \bar{\epsilon}}\right) q_0 \left(e^{i \Omega_- t} + e^{-i \Omega_+ t} \right) + q_0^\dagger \left(\frac{1 + i \sqrt{\bar{\epsilon}(2 + \bar{\epsilon})}}{1 + \bar{\epsilon}}\right) \left(e^{-i \Omega_- t} + e^{i \Omega_+ t} \right) \\
    p_a(t) &= 0 \\
    p_b(t) &= 0
\end{split}
\end{equation}
where $q_0$ is a complex constant with a magnitude determined by saturation effects in the sensor and independent of $\bar{\epsilon}$. 
The sensor's mean output photon flux is then given by photons in its amplitude quadrature. Using $q_\text{out} = \sqrt{2\gamma_a} q_a$ to relate the output amplitude quadrature to the amplitude quadrature in mode $\hat{a}$, and following the same analysis as in the phase-insensitive case of \cref{sec:classicalEOM}, we find that the phase-sensitive sensor's output photon flux is
\begin{equation}
    N_\pm = 4 \gamma_a |q_0|^2.
\end{equation}

\subsubsection{Linearized Dynamics}

After linearizing \cref{eq:phSensCoupledModeFull}, we find that the system's fluctuation dynamics in the rotating frame are given by
\begin{equation}\label{eq:phSensCoupledMode}
\begin{split}
    \partial_{t} \delta \hat{A} &= - \gamma_a \delta \hat{A} + \gamma_a (1 + \bar{\epsilon}) \delta \hat{B} + \gamma_a B \delta \epsilon + \sqrt{2 \gamma_a}\, \delta\hat{A}_\text{in} \\
    \partial_{t} \delta \hat{B} &= (\gamma_a - r)\delta \hat{B} + r \delta \hat{B}^\dagger - \gamma_a (1 + \bar{\epsilon} ) \delta \hat{A} - \gamma_a A \delta \epsilon + \sqrt{2 (\gamma_a + \gamma_b - r)}\, \delta\hat{B}_\text{amp}^\dagger + \sqrt{2 \gamma_b} \, \delta\hat{B}_\text{in}
\end{split}
\end{equation}
Using the mean dynamics of \cref{eq:phsensMeanDynamics}, we find that the system's quadrature fluctuations are governed by
\begin{equation}\label{eq:phsensLinearizedQuadDynamics}
\begin{split}
    \partial_{t} \delta \hat{q}_A =& - \gamma_a \delta \hat{q}_A + \gamma_a (1 + \bar{\epsilon}) \delta \hat{q}_B + \gamma_a \left(q_0^B (e^{i\Omega_- t} + e^{-i\Omega_+t}) + q_0^{B\dagger} (e^{-i\Omega_- t} + e^{i\Omega_+t}) \right) \delta \epsilon + \sqrt{2 \gamma_a}\, \delta\hat{q}_{A,\text{in}} \\
    \partial_{t} \delta \hat{q}_B =& \gamma_a \delta \hat{q}_B - \gamma_a (1 + \bar{\epsilon} ) \delta \hat{q}_A - \gamma_a \left(q_0 (e^{i\Omega_- t} + e^{-i\Omega_+t}) + q_0^{\dagger} (e^{-i\Omega_- t} + e^{i\Omega_+t}) \right) \delta \epsilon + \sqrt{2 (\gamma_a + \gamma_b - r)}\, \delta\hat{q}_{B,\text{amp}} \\
    &+ \sqrt{2 \gamma_b} \, \delta\hat{q}_{B,\text{in}} \\
    \\
    \partial_{t} \delta \hat{p}_A =& - \gamma_a \delta \hat{p}_A + \gamma_a (1 + \bar{\epsilon}) \delta \hat{p}_B + \sqrt{2 \gamma_a}\, \delta\hat{p}_{A,\text{in}}\\
    \partial_{t} \delta \hat{p}_B =& (\gamma_a - 2r) \delta \hat{p}_B - \gamma_a (1 + \bar{\epsilon} ) \delta \hat{p}_A  - \sqrt{2 (\gamma_a + \gamma_b - r)}\, \delta\hat{p}_{B,\text{amp}} + \sqrt{2 \gamma_b} \, \delta\hat{p}_{B,\text{in}},
\end{split}
\end{equation}
where $\hat{q}_{A,x}$ is the operator $\hat{q}_{a,x}$ moved into the rotating frame, and similarly for $\hat{p}_{A,x}$, $\hat{q}_{B,x}$, and $\hat{p}_{B,x}$ and we defined the complex constant $q_0^B \equiv q_0 (1-i\sqrt{\bar{\epsilon}(2+\bar{\epsilon})})/(1+\bar{\epsilon})$. Unlike the phase-insensitive sensor, the phase-quadrature equations contain no $\delta \epsilon$ terms since $p_a = p_b = 0$ for the phase-sensitive oscillator.

\subsubsection{Parameter Estimation}

Having derived the mean dynamics of the phase-sensitive EP sensor, we will now analyze its performance in estimating the mean value of $\epsilon$, $\bar{\epsilon}$, and in measuring a weak force coupled to $\delta\epsilon$. To evaluate its performance in estimating the mean value of $\bar{\epsilon}$, we take $\delta\epsilon = 0$ in \cref{eq:phSensCoupledMode} and consider the sensor's output phase quadrature spectrum.

First, we analyze the Fourier transform \cref{eq:phsensMeanDynamics} with $\delta \epsilon = 0$. Since the quadratures decouple, we can focus on the phase quadrature evolution given in the frequency domain by
\begin{equation}\label{eq:ohSensCoupledModeNoFluct}
\begin{split}
    (-i \omega + \gamma_a) \hat{p}_A [\omega] &=  \gamma_a (1 + \bar{\epsilon}) \delta \hat{p}_B [\omega] + \sqrt{2 \gamma_a}\, \delta\hat{p}_{A,\text{in}}[\omega] \\
    (-i \omega - \gamma_a + 2r) \delta \hat{p}_B [\omega] &= - \gamma_a (1 + \bar{\epsilon} ) \delta \hat{p}_A [\omega]  - \sqrt{2 (\gamma_a + \gamma_b - r)}\, \delta\hat{p}_{B,\text{amp}} [\omega] + \sqrt{2 \gamma_b} \, \delta\hat{p}_{B,\text{in}} [\omega]
\end{split}
\end{equation}
The sensor's output phase quadrature is related to the resonant mode quadrature operators of \cref{eq:ohSensCoupledModeNoFluct} by the boundary condition $\hat{p}_\t{out} = \sqrt{2 \gamma_a} \hat{p}_A - \hat{p}_{A,\t{in}}$, so the output phase quadrature is given in the rotating frame and the frequency domain by
\begin{equation}\label{eq:inOutPhSensPhaseQuad}
\begin{split}
    \delta p_{A,\text{out}} =& \frac{2(1+\bar{\epsilon})\gamma_a^{3/2}\sqrt{\gamma_a+\gamma_b-r}}{(\omega - \Omega_-)(\omega - \Omega_+)+2r(i\omega - \gamma)}\delta p_{B,\text{amp}} \\ 
    &+ \left( \frac{2\gamma_a(\gamma_a -2r + i\omega )}{(\omega - \Omega_-)(\omega - \Omega_+)+2r(i\omega - \gamma)} -1 \right) \delta p_{A,\text{in}} \\ &- \frac{2 (1+\bar{\epsilon})\gamma_a^{3/2}\sqrt{\gamma_b}}{(\omega - \Omega_-)(\omega - \Omega_+)+2r(i\omega - \gamma)} \delta p_{B,l},
\end{split}
\end{equation}
from which we can read off the transfer functions in the abstracted relation
\begin{equation}\label{eq:generalPhsensPhaseTfDef}
    \delta p_{A,\text{out}}[\omega] = H_\text{amp}^p[\omega] \delta p_{B,\text{amp}}[\omega] + H_\text{A,in}^p[\omega] \delta p_{A,\text{in}} [\omega] + H_\text{B.in}^p [\omega] \delta p_{B,\text{in}} [\omega].
\end{equation}

Using \cref{eq:inOutPhSensPhaseQuad,eq:generalPhsensPhaseTfDef} and assuming the noise modes are uncorrelated, we can write the output quadrature spectra in terms of the quadrature spectra of the noise operators as
\begin{equation}
    \bar{S}_{pp}^\text{out}[\omega] = \left|H_\text{A,in}^p [\omega]\right|^2 \bar{S}_{pp}^\text{A,in} [\omega] + \left|H_\text{amp}^p [\omega]\right|^2 \bar{S}_{pp}^\text{amp} [\omega] + \left|H_\text{B,in}^p [\omega]\right|^2 \bar{S}_{pp}^\text{B,in} [\omega].
\end{equation}
From this equation, we can calculate the output phase quadrature spectrum from this noise quadrature spectra, which we assume are those of thermal states with mean photon occupation number $\bar{n}_\t{th}$.
For nonzero losses, $\gamma_b>0$, the output phase quadrature spectrum is given to leading order in $\Delta \omega_\pm$ and $\bar{\epsilon}$ by
\begin{equation}\label{eq:SppPhSensLossy}
    \bar{S}_{pp}^\text{out}[\omega] = \frac{2\gamma_a^2 + 2\gamma_a \gamma_b - 3\gamma_a r + r^2}{2r^2}(1+2\bar{n}_\text{th}),
\end{equation}
which is independent of both $\bar{\epsilon}$ and $\Delta \omega_\pm$.

For the ideal lossless case with an amplifier that is not perfectly phase-sensitive, $r<\gamma_a$ , the output phase-quadrature spectrum is given by \cref{eq:SppPhSensLossy} with $\gamma_b=0$. However, for a purely phase-sensitive amplifier and no losses, $r=\gamma_a$ and $\gamma_b=0$, the expression on the right hand side of \cref{eq:SppPhSensLossy} is zero and the leading order output phase-quadrature spectrum is given instead by
\begin{equation}\label{eq:SppPhSensLossless}
    \bar{S}_{pp}^\text{out}[\omega] = \frac{\bar{\epsilon}(2+\bar{\epsilon}) (1 + 2 \bar{n}_\text{th})\Delta \omega_\pm^2}{2\gamma_a^2 (1 + \bar{\epsilon})^2},
\end{equation}
however, this result is not robust against loss, since the leading-order behavior is given by \cref{eq:SppPhSensLossy} for $\gamma_b > 0$.

\subsubsection{Weak Force Sensing}\label{sec:phSensWeakForce}

We have seen that a phase-sensitive EP sensor can have a sensing advantage in parameter estimation. To evaluate its advantage in weak force sensing, we consider \cref{eq:phsensLinearizedQuadDynamics}. First, considering the phase-quadrature component of this equation, we see that fluctuations in $\epsilon$ do not couple to the sensor's phase quadrature since it has no mean phase quadrature field. Fluctuations in $\epsilon$ are coupled to the sensor's amplitude quadrature, but, by design, the phase-sensitive EP sensor's amplitude quadrature dynamics are the same as those of a phase-insensitive EP sensor. Thus, like an EP sensor based around a phase-insensitive amplifier, the phase-sensitive EPsensor has no fundamental advantage in weak force sensing.

\subsection{Non-Markovian Dynamics}

As with the phase-insensitive PT-symmetric sensor, we can also evaluate the performance of a PT-symmetric sensor with a phase-sensitive amplifier in the non-Markovian regime. We do so by considering the sensor depicted in \cref{fig:fbOscEp} where the amplifier is now phase-sensitive. This sensor is described by the coupled mode equations
\begin{equation}\label{eq:loopEqEpPhSens}
\begin{split}
    \hat{a}_0 [\omega] &= \sqrt{\eta} \, \hat{a}_\text{in} [\omega] - \sqrt{1 - \eta} \, \hat{a}_\tau [\omega] \\
    \hat{a}_\text{out} [\omega] &= \sqrt{\eta} \, \hat{a}_\tau [\omega] + \sqrt{1 - \eta} \, \hat{a}_\text{in} [\omega] \\
    \hat{b} [\omega] &= \sqrt{\mu} \, \hat{a}_0 [\omega] + \sqrt{1 - \mu} \, \hat{b}_\text{s} [\omega] \\
    \hat{a} [\omega] &= \sqrt{\mu} \, \hat{b}_\text{s} [\omega] - \sqrt{1 - \mu} \, \hat{a}_0 [\omega] \\
    \hat{a}_\tau [\omega] &= e^{i \omega \tau} \hat{a} [\omega] \\
    \hat{b}_\tau [\omega] &= e^{i \omega \tau} \hat{b} [\omega] \\
    \hat{b}_\text{A} [\omega] &= \sqrt{1 + G} \, \hat{b}_\tau [\omega] + \sqrt{G}\, \hat{b}_\text{amp}^\dagger [\omega]\\
    \hat{b}_\text{s}[\omega] &= \cosh(\xi)\hat{b}_\text{A}[\omega] + \sinh(\xi)\hat{b}_\text{A}^\dagger[\omega]
\end{split}
\end{equation}
In writing the last two lines of \cref{eq:loopEqEpPhSens}, we have followed ref. \cite{LouSud23} in decomposing an arbitrary phase-sensitive amplifier as a phase-insensitive amplifier followed by an ideal single-mode squeezer with squeezing parameter $\xi \in \mathbb{R}$. 

\Cref{eq:loopEqEpPhSens} is most naturally analyzed in the quadrature basis. In the basis of raising and lowering operators, the last line of \cref{eq:loopEqEpPhSens} couples the lowering operators and their adjoints, giving a set of sixteen coupled equations. By switching to the quadrature basis, these equations decompose into two sets of eight coupled equations. For the amplitude quadrature, we have
\begin{equation}\label{eq:loopEqEpPhSensQ}
\begin{split}
    \hat{q}_0 [\omega] &= \sqrt{\eta} \, \hat{q}_\text{in} [\omega] - \sqrt{1 - \eta} \, \hat{q}_\tau^a [\omega] \\
    \hat{q}_\text{out} [\omega] &= \sqrt{\eta} \, \hat{q}_\tau^a [\omega] + \sqrt{1 - \eta} \, \hat{q}_\text{in} [\omega] \\
    \hat{q}_b [\omega] &= \sqrt{\mu} \, \hat{q}_0 [\omega] + \sqrt{1 - \mu} \, \hat{q}_\text{s} [\omega] \\
    \hat{q}_a [\omega] &= \sqrt{\mu} \, \hat{q}_\text{s} [\omega] - \sqrt{1 - \mu} \, \hat{q}_0 [\omega] \\
    \hat{q}_\tau^a [\omega] &= e^{i \omega \tau} \hat{q}_a [\omega] \\
    \hat{q}_\tau^b [\omega] &= e^{i \omega \tau} \hat{q}_b [\omega] \\
    \hat{q}_\text{A} [\omega] &= \sqrt{1 + G} \, \hat{q}_\tau^b [\omega] + \sqrt{G}\, \hat{q}_\text{amp}^\dagger [\omega]\\
    \hat{q}_\text{s}[\omega] &= e^{\xi} \hat{q}_\text{A}[\omega],
\end{split}
\end{equation}
and for the phase quadrature we have
\begin{equation}\label{eq:loopEqEpPhSensP}
\begin{split}
    \hat{p}_0 [\omega] &= \sqrt{\eta} \, \hat{p}_\text{in} [\omega] - \sqrt{1 - \eta} \, \hat{p}_\tau^a [\omega] \\
    \hat{p}_\text{out} [\omega] &= \sqrt{\eta} \, \hat{p}_\tau^a [\omega] + \sqrt{1 - \eta} \, \hat{p}_\text{in} [\omega] \\
    \hat{p}_b [\omega] &= \sqrt{\mu} \, \hat{p}_0 [\omega] + \sqrt{1 - \mu} \, \hat{p}_\text{s} [\omega] \\
    \hat{p}_a [\omega] &= \sqrt{\mu} \, \hat{p}_\text{s} [\omega] - \sqrt{1 - \mu} \, \hat{p}_0 [\omega] \\
    \hat{p}_\tau^a [\omega] &= e^{i \omega \tau} \hat{p}_a [\omega] \\
    \hat{p}_\tau^b [\omega] &= e^{i \omega \tau} \hat{p}_b [\omega] \\
    \hat{p}_\text{A} [\omega] &= \sqrt{1 + G} \, \hat{p}_\tau^b [\omega] - \sqrt{G}\, \hat{p}_\text{amp}^\dagger [\omega]\\
    \hat{p}_\text{s}[\omega] &= e^{-\xi} \hat{p}_\text{A}[\omega].
\end{split}
\end{equation}
We take the value of $\mu$ in \cref{eq:loopEqEpPhSensQ,eq:loopEqEpPhSensP} to be that given by \cref{eq:GandMuEP}, and we take $e^{\xi}\sqrt{1+G}\sqrt{1-\eta}=1$ such that the system has gain-loss balance for the amplitude quadrature. The sensor becomes purely phase sensitive when $G=0$ and $\xi = -\frac{1}{2}\ln (1-\eta)$. For the remainder of this section, we will use this equation to eliminate $G$ in favor of $\xi$. After taking these values for $G$ and $\mu$, we now use \cref{eq:loopEqEpPhSensQ,eq:loopEqEpPhSensP} to compute the transfer functions from the input and amplifier modes to the output in the non-Markovian PT-symmetric sensor with a phase-sensitive amplifier. We find
\begin{equation}\label{eq:phSensTfsNonMark}
\begin{split}
    H_\text{in}^q[\omega] &= \frac{e^{-i\omega\tau}\left[2+(2-\eta)e^{2 i \omega \tau}-3\eta+\eta^2-4 e^{i\omega\tau}(1-\eta)\cos\left(\frac{\eta}{2}\sqrt{\epsilon(2+\epsilon)}\right) \right]}{2 \sqrt{1-\eta}(2-\eta)\left[\cos(\omega\tau) - \cos\left(\frac{\eta}{2}\sqrt{\epsilon(2+\epsilon)}\right) \right]}\\
    H_\text{amp}^q[\omega] &= \frac{e^\xi\sqrt{\eta \left(e^{-2\xi}-1+\eta\right)\left[ 2 - \eta(2-\eta)-2(1-\eta)\cos\left(\eta\sqrt{\epsilon(2+\epsilon)}\right) \right]}}{2 \sqrt{1-\eta}(2-\eta)\left[\cos(\omega\tau) - \cos\left(\frac{\eta}{2}\sqrt{\epsilon(2+\epsilon)}\right) \right]}\\
    H_\text{in}^p[\omega] &= \frac{\sqrt{\eta\left( e^{-2\xi}+\eta-1 \right)\left[2-\eta(2-\eta)-2(1-\eta)\cos\left(\eta\sqrt{\epsilon(2+\epsilon)}\right) \right]}}{2\sqrt{1-\eta}\left[ (2-\eta)\cosh\left( \xi - i \frac{\eta}{2}\sqrt{\epsilon(2+\epsilon)} \right) + \eta \cos \left( \frac{\eta}{2}\sqrt{\epsilon(2+\epsilon)} \right) \sinh(\xi) -(2-\eta)\cos\left( \frac{\eta}{2}\sqrt{\epsilon(2+\epsilon)} \right) \cosh(\xi) \right]}\\
    H_\text{amp}^p[\omega] &= \frac{-4 (1-\eta)\cos \left( \frac{\eta}{2}\sqrt{\epsilon(2+\epsilon)}\right)\cosh(\xi)+ (2-\eta)\left[ (2-\eta)\cosh\left(\xi - i \frac{\eta}{2}\sqrt{\epsilon(2+\epsilon)} - \eta \sinh \left(\xi -i \frac{\eta}{2}\sqrt{\epsilon(2+\epsilon)} \right)\right) \right]}{2\sqrt{1-\eta}\left[ (2-\eta)\cosh\left( \xi - i \frac{\eta}{2}\sqrt{\epsilon(2+\epsilon)} \right) + \eta \cos \left( \frac{\eta}{2}\sqrt{\epsilon(2+\epsilon)} \right) \sinh(\xi) -(2-\eta)\cos\left( \frac{\eta}{2}\sqrt{\epsilon(2+\epsilon)} \right) \cosh(\xi) \right]}
\end{split}
\end{equation}

Comparing \cref{eq:phSensTfsNonMark} to \cref{eq:tfs}, we find that $H_\text{in}^q[\omega] = H_\text{in}[\omega]$ and $H_\text{amp}^q[\omega] = \sqrt{\frac{1}{\eta}-e^{2\xi}\left(\frac{1}{\eta}-1 \right)}\, H_\text{amp}[\omega]$, so the output amplitude quadrature of the PT-symmetric sensor with a phase-sensitive amplifier has the same spectral shape as the sensor with a phase-insensitive amplifier, as in the Markovian case. We also see that, as in the Markovian case, the output phase quadrature spectrum has a fundamentally different spectral shape than for the sensor with a phase-sensitive amplifier.

Using \cref{eq:phSensTfsNonMark}, and assuming the sensor's in-coupled mode and amplifier are in thermal states, the sensor's output phase-quadrature spectrum is given by
\begin{equation}\label{eq:SppOutPhSensNonM}
    \bar{S}_{pp}^\text{out}[\omega] = \left(\left| H_\text{in}^p[\omega] \right|^2 + \left| H_\text{in}^p[\omega] \right|^2 \right) \left(\frac{1}{2}+\bar{n}_\text{th} \right).
\end{equation}
We now consider \cref{eq:SppOutPhSensNonM} near the resonance frequencies and the exceptional point. For frequencies near resonance, $\omega = \Omega_0 \pm (\eta/(2 \tau))\sqrt{\epsilon (2 + \epsilon)} + \Delta \omega_\pm = \Omega_\pm + \Delta \omega_\pm$, the sensor's output phase quadrature spectrum is given by
\begin{equation}
    \bar{S}_{pp}^\text{out}[\omega] = \frac{\eta \left[ e^{-2 \xi} + \eta - 1 \right]\text{csch}(\xi)^2 + \left[ 2 - \eta(1 + \coth(\xi)) \right]^2}{8(1-\eta)}(1+2 \bar{n}_\text{th}),
\end{equation}
which reduces to \cref{eq:SppPhSensLossy}, with $\gamma_b = 0$, in the Markovian limit upon identifying the rate of phase-sensitive amplification, $r$ with $\xi/\tau$ and taking the limit that $\xi,\eta,\tau \rightarrow 0$ whith $\xi/\tau = r$ and $\eta / (2 \tau)$ held constant.
As with the Markovian PT-symmetric sensor with a phase-sensitive amplifier, this leading order term vanishes if the sensor uses a purely phase-sensitive amplifier, in which case the leading order term is
\begin{equation}
    \bar{S}_{pp}^\text{out}[\omega] = \frac{(1-\eta)(2-\eta)^2\epsilon \tau^2 \Delta \omega_\pm^2}{\eta^2}(1 + 2 \bar{n}_\text{th}),
\end{equation}
which reduces to \cref{eq:SppPhSensLossless} in the Markovian limit and the limit that $\epsilon \ll 1$, i.e. we operate the sensor near the PT-symmetric EP.

We see that as with PT-symmetric EP sensors using phase-insensitive amplifiers, those using phase-sensitive amplifiers have an order-unity correction to their Markovian spectra in the non-Markovian case.

\bibliographystyle{apsrev4-2}
\bibliography{ref_ep_quantum_noise}

\end{document}